\documentclass[10pt]{article}
\def\lsim{\mathrel{\lower2.5pt\vbox{\lineskip=0pt\baselineskip=0pt
          \hbox{$<$}\hbox{$\sim$}}}}
\def\gsim{\mathrel{\lower2.5pt\vbox{\lineskip=0pt\baselineskip=0pt
          \hbox{$>$}\hbox{$\sim$}}}}
\def\real{\mathrel{\lower.0pt \hbox{$I\!\!R$}}}
\renewcommand{\theequation}{\arabic{section}.\arabic{equation}}
\begin{document}   
%------------------------------------------------------------------------------
\markright{Braneworld gravity: Influence of the moduli fields \hfil}
%------------------------------------------------------------------------------
\def\Barcelo{Barcel\'o}
%------------------------------------------------------------------------------
\title{\Large 
\bf Braneworld gravity: Influence of the moduli fields}
\author{Carlos {\Barcelo} and Matt Visser\\[2mm]
{\small \it 
Physics Department, Washington University, 
Saint~Louis, Missouri 63130-4899, USA.}}
\date{{\small 4 September 2000; \LaTeX-ed \today}}
%------------------------------------------------------------------------------
\maketitle
%------------------------------------------------------------------------------
\begin{abstract}
We consider the case of a generic braneworld geometry in the presence
of one or more moduli fields ({\emph{e.g.}}, the dilaton) that vary
throughout the bulk spacetime. Working in an arbitrary conformal
frame, using the generalized junction conditions of gr-qc/0008008 and
the Gauss--Codazzi equations, we derive the effective ``induced''
on-brane gravitational equations.  As usual in braneworld scenarios,
these equations do not form a closed system in that the bulk can
exchange both information and stress-energy with the braneworld.  We
work with an arbitrary number of moduli fields described by an
arbitrary sigma model, with arbitrary curvature couplings, arbitrary
self interactions, and arbitrary dimension for the bulk. (The
braneworld is always codimension one.) Among the novelties we
encounter are modifications of the on-brane stress-energy conservation
law, anomalous couplings between on-brane gravity and the trace of the
on-brane stress-energy tensor, and additional possibilities for
modifying the on-brane effective cosmological constant. After
obtaining the general stress-energy ``conservation'' law and the
``induced Einstein equations'' we particularize the discussion to two
particularly attractive cases: for a ($n$--2)-brane in ([$n$--1]+1)
dimensions we discuss both the effect of (1) generic variable moduli
fields in the Einstein frame, and (2) the effect of a varying dilaton
in the string frame.
\vspace*{5mm}

\vspace*{5mm}

\noindent
V2 (5 Oct 2000): three references added, some references updated, no physics changes.
\end{abstract}
%-----------------------------------------------------------------------
\vfill
%----------------------------------------------------------------------
\hrule
%-----------------------------------------------------------------------
\bigskip
%-----------------------------------------------------------------------
\centerline{\underline{E-mail:} {\sf carlos@hbar.wustl.edu}}
%-----------------------------------------------------------------------
\centerline{\underline{E-mail:} {\sf visser@kiwi.wustl.edu}}
%-----------------------------------------------------------------------
\bigskip
%-----------------------------------------------------------------------
\centerline{\underline{Homepage:} {\sf http://www.physics.wustl.edu/\~{}carlos}}
%-----------------------------------------------------------------------
\centerline{\underline{Homepage:} {\sf http://www.physics.wustl.edu/\~{}visser}}
%-----------------------------------------------------------------------
\bigskip
%-----------------------------------------------------------------------
\centerline{\underline{Archive:}
{\sf hep-th/0009032}}
%----------------------------------------------------------------------
\bigskip
\hrule
\clearpage
%---------------------------------------------------------------------
% User definitions
%--------------------------------------------------------------------
\def\Box{\nabla^2}
%---------------------------
\def\d{{\mathrm d}}
%----------------------------
\def\ie{{\em i.e.\/}}
\def\eg{{\em e.g.\/}}
\def\etc{{\em etc.\/}}
\def\etal{{\em et al.\/}}
%----------------------------
\def\S{{\mathcal S}}
\def\I{{\mathcal I}}
\def\L{{\mathcal L}}
\def\H{{\mathcal H}}
\def\M{{\mathcal M}}
\def\R{{\mathcal R}}
\def\K{{\mathcal K}}
\def\F{{\mathcal F}}
\def\E{{\mathcal E}}
%-----------------------------
\def\eff{{\mathrm{eff}}}
\def\Newton{{\mathrm{Newton}}}
\def\Einstein{{\mathrm{Einstein}}}
\def\bulk{{\mathrm{bulk}}}
\def\brane{{\mathrm{brane}}}
\def\matter{{\mathrm{matter}}}
\def\tr{{\mathrm{tr}}}
\def\normal{{\mathrm{normal}}}
\def\implies{\Rightarrow}
\def\half{{1\over2}}
\def\effective{{\mathrm{effective}}}
\def\anomalous{{\mathrm{anomalous}}}
\def\quadratic{{\mathrm{quadratic}}}
%-------------------------------------
\def\Eotvos{{E\"otv\"os}}
%-------------------------------------
\def\HRULE{\bigskip\hrule\bigskip}
%----------------------------------------------------------------------- 
%\def\SIZE{1.00} % for graphics package
%-----------------------------------------------------------------------

%------------------------------------------------------------------------------
\section{Introduction}
%------------------------------------------------------------------------------
\label{S:introduction}
%------------------------------------------------------------------------------

The idea that our observable universe might be a submanifold of a
higher-dimensional spacetime is an old one, going back at least 18
years~\cite{old-non-compact}. This idea has recently been revived and
extensively developed, leading to various versions of the
``braneworld'' scenario, with two major variants depending on whether
the extra dimensions are large but compact~\cite{large-compact} or
truly non-compact~\cite{RS,Gogberashvili,Arkani-et-al}.  In particular
a key issue is the form of the effective Einstein equations that are
induced on the brane by what amounts to a ``dimensional reduction''
procedure. Particularly important are the papers of Shiromizu, Maeda,
and Sasaki~\cite{Shiromizu}, Maeda and Wands~\cite{Wands}, and Mennim
and Battye~\cite{Mennim}. See also~\cite{background-1,background-2}.

In this paper we shall extend this approach and consider the case of a
braneworld geometry in the presence of one or more moduli fields (\eg,
the dilaton) that vary throughout the bulk spacetime. We generalize
the previous calculations by working with an arbitrary number of
moduli fields (instead of just the one dilaton field) described by an
arbitrary sigma model (instead of limiting attention to kinetic
energies that are canonical in at least one conformal frame), with
arbitrary curvature couplings (equivalent to working in arbitrary
conformal frame), arbitrary self interactions, and arbitrary dimension
for the bulk.  [While the most popular braneworld scenario involves
reduction from (4+1) to (3+1) dimensions as essentially the last step
in arriving at a phenomenologically acceptable model, we wish to leave
open the possibility of, for instance, performing several braneworld
reductions in sequential stages.  At each stage of the reduction the
braneworld is always codimension one in the corresponding bulk.]

Using the generalized junction conditions
of~\cite{generalized-junction}, which are the appropriate
generalization of the Israel--Lanczos--Sen junction
conditions~\cite{Israel-Lanczos-Sen} to arbitrary conformal frame, and
the Gauss--Codazzi equations we derive effective on-brane
gravitational equations.  As usual in braneworld scenarios, these
equations do not form a closed system in that the bulk can exchange
both information and stress-energy with the braneworld.  In
particular, the on-brane surface stress energy is not conserved in the
usual sense: braneworld stress-energy can both flow into and put of
the bulk, and in addition braneworld stress-energy can couple to
``along the brane'' variations in the moduli fields.  For the
``induced Einstein equations'' novelties include anomalous couplings
between braneworld gravity and the trace of the braneworld
stress-energy tensor, and additional possibilities for modifying the
on-brane effective cosmological constant.

After obtaining general ``induced Einstein equations'' we
particularize the discussion to a two particularly attractive cases:
for a generic ($n$--2)-brane in ([$n$--1]+1) dimensions we discuss (1)
generic variable moduli fields in the Einstein frame, and (2) a
variable dilaton field in the string frame.

%------------------------------------------------------------------------
\section{Lagrangian and generalized junction conditions}
%------------------------------------------------------------------------
\setcounter{equation}{0}
%---------------------------------------------------------------------

A key technical complication in the current calculation is the use of
a general conformal frame. This allows our formalism to be applied
equally well in the Einstein frame, Jordan frame, or string
frame. There is continuing debate as to which conformal frame is the
``most physical'', with our own attitude being that it depends on the
physical questions you are asking. A nice review, with additional
references is given by Faraoni, Gunzig, and
Nardone~\cite{conformal-frames}. We will not delve further into this
issue, and in this current paper keep the choice of conformal frame
arbitrary.

%------------------------------------------------------------------------
\subsection{General conformal frame}
%------------------------------------------------------------------------

We consider an action of the form
\begin{eqnarray}
\S=&&\hspace{-6mm}
{1 \over 2}\int_{{\rm int}(\M)} \sqrt{-g} \;\d^n x \; \kappa_n^{2} \;
F(\phi) \; \left[ R -2\Lambda \right] 
-\int_{\partial\M} \sqrt{-q} \;\d^{n-1}x \;  \kappa_n^{2} \; F(\phi)\;K
\nonumber\\  
&&\hspace{-6mm}
+\int_{{\rm int}(\M)} \sqrt{-g} \;\d^n x \;
\left\{ -{1 \over 2} 
H_{ij}(\phi) \; 
\left[ g^{AB} \; \partial_{A}\phi^i \; \partial_{B}\phi^j \right] 
-V(\phi,\psi) 
+\L_\bulk(g_{AB},\phi,\psi)
\right\}
\nonumber \\  
&&\hspace{-6mm}
+\int_{\rm brane} \sqrt{-q} \;\d^{n-1}x\;\L_\brane(q_{AB},\phi,\psi).
\label{E:action}
\end{eqnarray}
Here $\kappa_n$ has dimensions $(length)^{1-n/2}=(energy)^{n/2-1}$;
the same as the dimensions of the moduli field $\phi$; as usual we
choose units so that $\hbar=1$ and $c=1$.  (Warning: In most of the
extant literature $\kappa_n$ is defined as above, however several key
papers, such as~\cite{Shiromizu,Wands}, invert the definition of
$\kappa_n$.) We have slightly generalized the action
of~\cite{generalized-junction} by explicitly exhibiting the
$n$-dimensional Newton constant (encoded in $\kappa_n$), and allowing the bulk
Lagrangian to posses additional non-derivative dependence on the
moduli fields.  A tricky point is that the Gibbons--Hawking boundary
term now takes the form~\cite{generalized-junction}
\begin{equation}
-\int_{\partial\M} \sqrt{-q} \;\d^{n-1}x \;  \kappa_n^{2} \; F(\phi)\;K.
\end{equation}
The three arbitrary functions, $F(\phi)$, $H_{ij}(\phi)$, and $V$,
allow us to deal with an extremely wide class of possible moduli
fields, one that covers essentially every possibility extant in the
literature.  To also permit interactions between the brane and the
bulk fields, in our analysis we allow the brane Lagrangian to depend
arbitrarily on the metric and on the bulk scalar fields, (this is in
addition to its dependence on the ``matter'' fields trapped on or near
the brane).

In the bulk region surrounding the brane, we adopt Gaussian normal
coordinates with $\eta$ denoting the spacelike normal to the brane,
and $q_{AB} = g_{AB} - n_A\; n_B$ denoting the induced metric.  The
generalized junction conditions derived in~\cite{generalized-junction}
read
\begin{eqnarray}
\pi^{+}_{AB}- \pi^{-}_{AB}
=
{1 \over 2}  \sqrt{-q}  \;S_{AB}, 
\label{E:ilsphi1}
\\ 
\pi^{+}_{\phi^i}-\pi^{-}_{\phi^i}
= 
\sqrt{-q} \; 
{\partial \L_\brane \over \partial \phi^i},
\label{E:ilsphi2}
\\ 
\pi^{+}_{\psi}-\pi^{-}_{\psi}
= 
\sqrt{-q} \; 
{\partial \L_\brane \over \partial \psi}.
\label{E:ilsphi3}
\end{eqnarray}
The gravitational ``momentum'' is defined
by~\cite{generalized-junction}
\begin{equation}
\pi_{AB}
=
\kappa_n^2 \left\{
{1 \over 2} \sqrt{-q} \; F(\phi)\; \left(K_{AB}-q_{AB}K\right)
+{1 \over 2} \sqrt{-q} \; F'_i(\phi)\; 
\left({\partial\phi^i\over\partial\eta}\right) q_{AB}
\right\}.
\label{E:pigphi} 
\end{equation}
(Here $\eta$, although spacelike, is for technical reasons formally
treated as though it were an evolution parameter.)  For convenience we
have defined
\begin{equation}
 F'_i(\phi) \equiv \partial_i F(\phi) 
\equiv {\partial F(\phi)\over\partial\phi^i}.
\end{equation}
The ``momentum'' canonically conjugate to $\phi$
is~\cite{generalized-junction}
\begin{equation}
\pi_{\phi^i}
=
-\sqrt{-q}\;H_{ij}(\phi)\; \left({\partial\phi^j\over\partial\eta}\right) 
-\sqrt{-q}\; \kappa_n^2 \;F'_i(\phi)\; K.
\label{E:piphi}
\end{equation}
While there are additional junction conditions for the ``matter''
fields $\psi$, they are not germane to the present discussion ---
see~\cite{generalized-junction} for details.  Rearranging the previous
expressions [(\ref{E:ilsphi1}) to (\ref{E:piphi})] and defining the
discontinuities
\begin{equation}
\K_{AB} \equiv K_{AB}^+ - K_{AB}^-, 
\qquad \qquad 
J^i = {\partial\phi^i\over\partial n}^+ -  {\partial\phi^i\over\partial n}^-,
\end{equation}
we arrive at
\begin{eqnarray}
&&
F \; \left(\K_{AB} - q_{AB} \; \K\right) 
+ F'_i \; J^i \; q_{AB} 
=  \kappa_n^{-2}  \;S_{AB},
\label{E:j1}
\\
\nonumber
\\
&&
-H_{ij} \; J^j - \kappa_n^2 \; F'_i \; \K = 
{\partial \L_\brane \over \partial \phi^i}.
\label{E:j2}
\end{eqnarray}
We have found that it is extremely useful to separate equation
(\ref{E:j1}) into a trace-free portion and a trace. That is
\begin{eqnarray} 
&&
\K_{AB} - {1\over n-1} \; \K \; q_{ab} =
{\kappa_n^{-2}\over F} 
\; \left(S_{AB} - {1\over n-1} \; S \; q_{ab}\right), 
\label{E:j-tracefree}
\\
\nonumber
\\
&&
(n-2) \; F \; \K - (n-1) \; F'_i \; J^i =  -  \kappa_n^{-2} \;S.
\label{E:j3}
\end{eqnarray}
Inverting the general equations (\ref{E:j2}) and (\ref{E:j3}) yields
the generalized junction conditions.  For the discontinuity in the
trace of extrinsic curvature
\begin{equation}
\K=
-{
 \kappa_n^{-2} \; S + (n-1) H^{ij} \; F'_i \;(\L_\brane)'_j
\over
(n-2) F + (n-1) H^{kl} \;\kappa_n^2 \; F'_k \; F'_l
}.
\label{E:j-trace}
\end{equation}
For the normal discontinuity in the scalar derivative
\begin{eqnarray} 
J^i &=& 
H^{ij}  
\left[ 
F'_j \; 
{ 
S + (n-1) H^{pq} \; \kappa_n^{2} \; F'_p  \;(\L_\brane)'_q 
\over 
(n-2) F + 
(n-1) H^{kl} \;  \kappa_n^2 \; F'_k \; F'_l
}
-  (\L_\brane)'_j 
\right]
\\
&=&   
{
H^{ij}\;  F'_j \; 
\over
(n-2) F + (n-1) H^{kl} \; \kappa_n^2 \; F'_k \; F'_l 
} \; S
\\
&&-
\left[ H^{ij}  -
{
(n-1) \kappa_n^2 \; H^{ip} \; F'_p  \; H^{jq}  \; F'_q
\over
(n-2) F + (n-1) H^{kl} \; \kappa_n^2\; F'_k \; F'_l 
}
\right]
\;(\L_\brane)'_j. 
\nonumber
\label{E:j-phi} 
\end{eqnarray} 
Reassembling the trace and trace-free parts of the extrinsic curvature
\begin{equation} 
\K_{AB} 
= 
\kappa_n^{-2}\; {S_{AB} \over F } 
- 
\left\{ 
{ 
[F + H^{ij} \;\kappa_n^2 \; F'_i \; F'_j ] \kappa_n^{-2}\; S  
+ F \; H^{ij}\; F'_i\; (\L_\brane)'_j 
\over 
(n-2) F +  (n-1) H^{ij}\;  \kappa_n^2\; F'_i\;F'_j
}  
\right\} 
\; {q_{AB}\over F}. 
\label{E:j-reassemble} 
\end{equation} 
These expressions are rather unwieldy and for computations we have
found it useful to introduce dimensionless coefficients
$\gamma_{ij}(\phi)$ according to the scheme
\begin{eqnarray}
\K &=& {\kappa_n^{-2} \over F} 
\left( 
\gamma_{11} \; S + \kappa_n \; \gamma_{12}^i \; (\L_\brane)'_i
\right),
\label{E:j-k-2}
\\
J^i &=& {\kappa_n^{-1} \over F} 
\left( 
\gamma_{21}^i \; S + \kappa_n \;  \gamma_{22}^{ij} \; (\L_\brane)'_j
\right), 
\label{E:j-phi-2}
\\
\K_{AB}
&=&
{ \kappa_n^{-2} \over F } \; 
\left(
S_{AB} + \tilde\gamma_{11} \; S \; q_{AB} + 
\kappa_n \; \tilde\gamma_{12}^i \; (\L_\brane)'_i \; q_{AB}
\right).
\label{E:j-reassemble-2}
\end{eqnarray}
These dimensionless coefficients depend only on the value of the
moduli fields on the brane itself (and the dimensionality of
spacetime). Explicit formulae are given in Appendix A.

Insofar as the junction conditions (and even the bulk Einstein
equations) are concerned, there are definite advantages to performing
a conformal redefinition of fields and going to the Einstein frame
$F_E(\phi)=1$.  However going to the Einstein frame usually carries a
cost that causes problems elsewhere in the analysis: For instance the
Einstein frame is not the appropriate frame for asking questions about
string propagation (the string frame is better adapted to that), while
in Brans--Dicke theories (and their relatives) the Einstein frame is
inappropriate for discussing \Eotvos-type experiments (universality of
free fall; the Jordan frame is more appropriate for that). This is why
we are keeping the choice of conformal frame arbitrary.

%------------------------------------------------------------------------
\subsection{Einstein frame}
%------------------------------------------------------------------------

To make this a little more explicit, suppose we start with the general
Lagrangian above and {\emph{define}} the Einstein frame metric by
\begin{equation}
[g_E]_{AB}
= 
F(\phi)^{2/(n-2)} \; g_{AB}; 
\qquad
[q_E]_{AB}
= 
F(\phi)^{2/(n-2)} \; q_{AB}. 
\end{equation}
Then it is a standard computation to verify that
\begin{eqnarray}
&&{1\over2}\int_{{\rm int}(\M)} \sqrt{-g} \;\d^n x \; F(\phi) \;  R(g)
-\int_{\partial\M} \sqrt{-q} \;\d^{n-1}x \; F(\phi)\;K(g)
\nonumber\\
&&
\qquad 
=
{1\over2}\int_{{\rm int}(\M)} \sqrt{-g_E} \;\d^n x \;  R(g_E)
-\int_{\partial\M} \sqrt{-q_E} \;\d^{n-1}x \;K(g_E)
\nonumber\\
&&
\qquad\qquad
-{1\over2}\int_{{\rm int}(\M)} \sqrt{-g_E} \;\d^n x \; {n-1\over n-2} \;
[g_E]^{AB} \; {\partial_A F \; \partial_B F\over F^2}.
\end{eqnarray}

Then the action for our general theory expressed in the Einstein frame
becomes
\begin{eqnarray}
\S_\Einstein=&&\hspace{-6mm}
{1 \over 2}\int_{{\rm int}(\M)} \sqrt{-g_E} \;\d^n x \; \kappa_n^{2} \; 
\left[ R(g_E) -2 F(\phi)^{-2/(n-2)} \; \Lambda \right] 
\nonumber\\
&&
-\int_{\partial\M} \sqrt{-q_E} \;\d^{n-1}x \;  \kappa_n^{2} \;K(g_E)
\nonumber\\  
&&\hspace{-6mm}
+\int_{{\rm int}(\M)} \sqrt{-g_E} \;\d^n x \;
\Bigg\{ -{1 \over 2} 
[H_E]_{ij}(\phi) \; 
\left[ g^{AB} \; \partial_{A}\phi^i \; \partial_{B}\phi^j \right] 
\nonumber
\\
&&
\qquad
- F(\phi)^{-n/(n-2)} \; V(\phi,\psi) 
\nonumber
\\
&&
\qquad
+ F(\phi)^{-n/(n-2)} \; \L_\bulk(F(\phi),[g_E]_{AB},\phi,\psi)
\Bigg\}
\nonumber \\  
&&\hspace{-6mm}
+\int_{\rm brane} \sqrt{-q_E} \;\d^{n-1}x\; 
F(\phi)^{-(n-1)/(n-2)}\;\L_\brane(F(\phi),[q_E]_{AB},\phi,\psi).
\label{E:action-in-einstein-frame}
\end{eqnarray}
Here the Einstein-frame sigma model metric is
\begin{equation}
[H_E]_{ij}(\phi)  \equiv
{H_{ij}(\phi) \over F(\phi) } + 
\kappa_n^2 \; {n-1\over n-2} \; {F'_i(\phi)  \; F'_j (\phi) \over F^2(\phi) },
\label{E:sigma-model-metric}
\end{equation}
and we can reduce clutter by defining
\begin{equation}
V_E(\phi,\psi) \equiv F(\phi)^{-n/(n-2)} \; V(\phi,\psi),
\end{equation}
\begin{equation}
\L_\bulk^E([g_E]_{AB},\phi,\psi) \equiv 
F(\phi)^{-n/(n-2)} \; \L_\bulk(F(\phi),[g_E]_{AB},\phi,\psi),
\end{equation}
and
\begin{equation}
\L_\brane^E([q_E]_{AB},\phi,\psi) \equiv
F(\phi)^{-(n-1)/(n-2)}\;\L_\brane(F(\phi),[q_E]_{AB},\phi,\psi).
\end{equation}

The various prices that are paid for making the gravity sector look
simple include: (1) what was a simple bulk cosmological constant in
the original conformal frame has now become a moduli dependent
potential; (2) even if the kinetic energies are canonical in the
original conformal frame, there will generally be a non-trivial
sigma-model metric in the Einstein frame; (sometimes you can
ameliorate this by additionally redefining the moduli fields in a
frame-dependent manner); (3) there are now additional (implicitly
moduli-dependent) terms in both bulk and brane Lagrangians. If you are
willing to live with all this then the gravitational sector at least
is considerably simpler and the Einstein-frame junction conditions
read
\begin{equation}
[\K_E]_{AB}
=
\kappa_n^{-2} \; \left\{
[S_E]_{AB}-{S_E \over n-2 } \; [q_E]_{AB}
\right\},
\label{E:k-einstein-frame}
\end{equation}
and
\begin{equation}
[J_E]^i= - [H_E]^{ij} \;\; (\L_\brane^E)'_j.
\label{E:j-phi-einstein-frame}
\end{equation}
Sometimes (but not always) the gain in the junction conditions is
worth the price paid elsewhere in the system. (There is a long and
contentious debate in the literature concerning which frame is the
most ``physical''; see~\cite{conformal-frames} for details and
references.)

%-----------------------------------------------------------------------
\section{Effective stress-energy tensor (general frame)}
%-----------------------------------------------------------------------
\label{S:effective}
%-----------------------------------------------------------------------
\setcounter{equation}{0}
%---------------------------------------------------------------------

For the general systems under consideration (\ref{E:action}) the
bulk Einstein equation reads
\begin{eqnarray}
G_{AB}&\equiv& R_{AB}-{1 \over 2}g_{AB}R
\nonumber\\
&=& 
\kappa_n^{-2} \; F^{-1}(\phi) \; H_{ij}(\phi)
\left[
\partial_{A}\phi^i \; \partial_{B}\phi^j 
-{1 \over 2} g_{AB} \;\;
g^{CD} \;\partial_{C}\phi^i \; \partial_{D}\phi^j
\right] 
\nonumber \\
&&
+F^{-1}(\phi) \; 
\left[
\nabla_A \nabla_B F(\phi)
-g_{AB}  \; \; g^{CD} \;\nabla_C \nabla_D F(\phi)
\right]
\nonumber \\
&&
+\kappa_n^{-2}\;F^{-1}(\phi) \; T_{AB}^\psi
- g_{AB} \;\kappa_n^{-2}\;F^{-1}(\phi)\; V(\phi,\psi)
- g_{AB} \;\Lambda.
\label{E:bulk-einstein-tensor}
\end{eqnarray}
In deriving this we have first varied with respect to the metric, and
then systematically rearranged terms until the equation is cast in the
form $G_{AB} = \kappa_n^{-2}\;T^\eff_{AB}$; see~\cite{scalars} for
some examples of this procedure.  The double derivatives acting on $F$
can be recast as
\begin{equation}
\nabla_A \nabla_B F(\phi) = 
F'_i(\phi) \; (\nabla_A \nabla_B \phi^i)  
+
F''_{ij}(\phi) \; (\nabla_A \phi^i) \; (\nabla_B \phi^j).
\end{equation}
This allows us to define the effective bulk stress-energy tensor as
\begin{eqnarray}
T_{AB}^\effective &=& 
F^{-1}(\phi) \; 
\left(H_{ij}(\phi) + \kappa_n^2 \; F''_{ij}(\phi) \right)
\partial_{A}\phi^i \; \partial_{B}\phi^j 
\nonumber \\
&& - 
F^{-1}(\phi)  \;  {1 \over 2} \;
\left(H_{ij}(\phi) + 2 \kappa_n^2 \; F''_{ij}(\phi) \right)
\; g_{AB} \;\;
g^{CD} \;\partial_{C}\phi^i \; \partial_{D}\phi^j 
\nonumber \\
&&
+F^{-1}(\phi) \;  F'_i(\phi) \; \kappa_n^2 \;
\left(
\nabla_A \nabla_B \phi^i - g^{CD} \;\nabla_C \nabla_D\phi^i
\right)
\nonumber \\
&&
+F^{-1}(\phi) \; T_{AB}^\psi
- g_{AB} \;F^{-1}(\phi)\; V(\phi,\psi)
- g_{AB}\; \kappa_n^2 \; \Lambda.
\label{E:bulk-einstein-tensor-2}
\end{eqnarray}
This is a generalization of the effective stress-energy for ordinary
non-minimally coupled scalars, which correspond to $H(\phi)=1$, and
$F(\phi)=1-\xi\kappa_n^{-2}\phi^2$. For details see~\cite{scalars}.
The bulk scalar field equation reads
\begin{eqnarray}
\nabla^A\left(H_{ij}(\phi)\;\nabla_A\phi^j\right)
&=&
V'_i(\phi) - (\L_\bulk)'_i - \half \kappa_n^2\;F'_i(\phi) [R-2\Lambda] 
\nonumber\\
&&
+ \half (\partial_i H_{jk}) \nabla_A\phi^j \; \nabla^A\phi^k.
\end{eqnarray}
There is an additional equation of motion for the generic ``matter''
fields that will not concern us.

%-----------------------------------------------------------------------
\section{The Gauss--Codazzi equations}
%-----------------------------------------------------------------------
\setcounter{equation}{0}
%---------------------------------------------------------------------

In order to see how effective ($n$--1)-dimensional gravity is induced
on the brane we follow~\cite{Shiromizu} and begin with the equations
of Gauss and Codazzi. We use them to relate the extrinsic curvature
and ($n$--1)-dimensional Riemann tensor of a hypersurface to the
distribution of bulk matter (this hypersurface being taken to coincide
with the location of the brane). The Codazzi equation is related to
the conservation of stress-energy on the brane
({\emph{non-conservation} in certain circumstances), while the Gauss
equation provides an ``induced'' generalized Einstein equation for the
braneworld.

%-----------------------------------------------------------------------
\subsection{The Codazzi equation and its implications}
%-----------------------------------------------------------------------

Consider the Codazzi equation~\cite{Shiromizu}
\begin{equation}
D^A (K_{AB}^\pm -  K^\pm q_{AB}) = - {}^{(n)}R_{AA'}^\pm \; n^A \; q^{A'}{}_B.
\end{equation}
Here $D$ (as opposed to $\nabla$) is the intrinsic covariant
derivative with respect to $q$ defined on the brane. We adopt
MTW~\cite{MTW} conventions regarding the sign of the extrinsic
curvature, which differs from that of~\cite{Shiromizu}. Note that in
general there are two Codazzi equations, one for each side of the
brane (unless one is adopting the ``one-sided'' approach
of~\cite{void,edge}).

First apply the bulk Einstein equation
\begin{equation}
D^A K_{AB}^\pm - D_B K^\pm = 
-\kappa_n^{-2} \; \; {}^{(n)}T_{AA'}^\pm \; n^A \; q^{A'}{}_B.
\end{equation}
Now adopt ``reduced'' Gaussian normal coordinates to simplify this
expression: Let indices such as $a$, $b$, $c$ run from $0$ to $n-2$,
so that they denote coordinates on (or parallel to) the brane. In
contrast, indices such as $A$, $B$, $C$ run from $0$ to $n-1$, and
denote coordinates in the bulk. The index $n$ will denote the
direction normal to the brane. Then
\begin{equation}
g_{AB} =
\left[
\begin{array}{cc}q_{ab} & 0\\ 0 & 1 \end{array}
\right];
\qquad
q_{AB} =
\left[
\begin{array}{cc}q_{ab} & 0\\ 0 & 0 \end{array}
\right].
\end{equation}
In these coordinates
\begin{equation}
K_{AB} =
\left[
\begin{array}{cc}K_{ab} & 0\\ 0 & 0 \end{array}
\right].
\end{equation}
It is useful to decompose the bulk stress energy tensor as
follows:
\begin{equation}
T_{AB} \equiv 
\left[
\begin{array}{cc}T_{ab} & f_a\\ f_b & T_{nn} \end{array}
\right].
\end{equation}
Here $T_{ab}$ denotes the in-brane part of the bulk stress-energy;
$f_a \equiv T_{an}$ denotes the flux onto (or away from) the
brane---this implies a shear force applied to the brane by the bulk
matter, and $T_{nn}$ denotes the normal compressive force (pressure)
on the brane. (These ``reduced'' on-brane coordinates are often much
easier to work with than the full $n$-dimensional system; however as
much of the literature uses the full system we shall aim for maximum
comprehensibility by keeping certain key formulas in the ``full''
system.)

In these ``reduced'' coordinates
\begin{equation}
D^a \left(K_{ab}^\pm - K^\pm\; q_{ab}\right) = -\kappa_n^{-2} \; f_b^\pm.
\end{equation}
Because the flux term (implying a shear force) this is not quite a
conservation equation.  Now take differences between the two sides of
the brane and define
\begin{equation}
\F_a \equiv f^+_a - f^-_a.
\end{equation}
This represents the discontinuity in the flux of bulk matter as one
crosses the brane---it describes how much net bulk matter is accreting
onto or evaporating from the brane. The resulting ``conservation law''
is
\begin{equation}
D^a \left(\K_{ab} - \K\; q_{ab}\right) = -\kappa_n^{-2} \; \F_b.
\end{equation}
Now apply the junction condition to write $\K_{AB}$ in terms of the
surface stress-energy tensor and the discontinuity of the normal
derivative of the moduli fields. In particular, write [using
(\ref{E:j1})]
\begin{eqnarray}
\K_{ab} - \K\; q_{ab} 
&=& 
{\kappa_n^{-2}\over F} S_{ab} - {F'_i \; J^i\over F} \; q_{ab}
\nonumber\\
&=&
{\kappa_n^{-2}\over F} S_{ab} 
- {F'_i \over F^2} \;  
\left( 
\kappa_n^{-1} \; \gamma_{21}^i \; S + \gamma_{22}^{ij} \; (\L_\brane)'_j
\right) \; q_{ab}
\nonumber\\ 
&=&
{\kappa_n^{-2}\over F} 
\left( S_{ab} + \alpha \; q_{ab} \;S \right)
+{\kappa_n^{-1}\over F} \; \beta^i \; (\L_\brane)'_i \; q_{ab}.
\end{eqnarray}
This serves as the definition of the dimensionless functions $\alpha$
and $\beta$ (which depend on the on-brane values of the moduli fields;
explicit formulae are given in Appendix A).  Then
\begin{equation}
D^a \left( 
{S_{ab} + \alpha\;q_{ab}\;S + \kappa_n \; \beta^i \; q_{ab} \; (\L_\brane)'_i \over F} 
\right) = 
-\F_b.
\end{equation}
Rearranging this
\begin{eqnarray}
D^a S_{ab} &=& 
-F \; \F_b 
+ 
{D^a F\over F}  
\left( 
S_{ab} + \alpha\;q_{ab}\;S + \kappa_n \beta^i \; (\L_\brane)'_i
\right)
\nonumber\\
&&-
D_b  \left( 
\alpha \;S + \kappa_n \beta^i \ (\L_\brane)'_i 
\right).
\end{eqnarray}
Therefore
\begin{eqnarray}
D^a S_{ab} &=& 
-F \; \F_b + 
{F'_i \; D^a \phi^i \over F}
\left( 
S_{ab} + \alpha\;q_{ab}\;S + \kappa_n \beta^i \; (\L_\brane)'_i
\right)
\nonumber\\
&&-
D_b \phi^j \left( 
\alpha'_j \;S + \kappa_n (\beta^i)'_j \; (\L_\brane)'_i
\right)
-
\left( 
\alpha \;D_b S + \kappa_n \beta^i \ D_b (\L_\brane)'_i 
\right).
\label{E:non-conservation}
\end{eqnarray}
In short, stress energy on the brane is {\emph{not}} conserved---the
nonconservation arises both from net bulk fluxes onto the brane, and
from variations in the value of the moduli fields as one moves along
the brane. If the moduli fields on the brane are ``translationally
invariant'' (as for instance in FLRW brane cosmologies before one
switches on perturbations) the second and third term on the RHS
vanish (since then $D_a \phi^i=0$).  Even if the moduli fields on
the brane are translationally invariant, the last term on the RHS can
still pick up contributions from $D_b S$ and $D_b (\L_\brane)'_j$;
these are best dealt with on a case by case basis.  In most models
considered to date one imposes $\F_a=0$ by symmetry or by fiat (see
\eg, \cite{Shiromizu}) and so recovers a conservation law for the ordinary
on-brane surface stress-energy---we now see that in general this is an
approximation.

Note that the analysis so far has been very general, we have not had
to impose either $Z_2$ symmetry or ``one-sidedness''. If one chooses
to work directly in the Einstein frame, then
\begin{equation}
D^a_E \left([\K_E]_{ab} - [\K_E] \; [q_E]_{ab} \right)= 
-\kappa_n^{-2}  \; [\F_E]_b.
\end{equation}
The moduli fields have now (apparently) decoupled from the
conservation law, but it is more correct to say that they have gone
underground by modifying the definition of $\F\to\F_E$. In terms of
the surface stress-energy we have the much simpler relation
\begin{equation}
D^a_E [S_E]_{ab} =  -[\F_E]_b.
\label{E:codazzi-in-einstein}
\end{equation}
%

%-----------------------------------------------------------------------
\subsection{The Gauss equation and its implications}
%-----------------------------------------------------------------------

In order to see how effective ($n$--1)-dimensional gravity is `induced
on the braneworld we begin with the Gauss equation and write the
($n$--1)-dimensional Riemann tensor of the braneworld in the
form~\cite{Shiromizu}
\begin{equation}
{}^{(n-1)}R_{ABCD}= 
{}^{(n)}R_{A'B'C'D'}^\pm \; \;
q^{A'}{}_A \;  q^{B'}{}_B \; q^{C'}{}_C \; q^{D'}{}_D \;
+ K_{AC}^\pm\; K_{BD}^\pm - K_{AD}^\pm \; K_{BC}^\pm.
\end{equation}
This equation applies to each side of the brane independently.  After
repeated contractions and some rearrangement\footnote{%
%-----------------------------------------------------------------
Comment: Equations (1) through (6) of the Shiromizu--Maeda--Sasaki
analysis~\cite{Shiromizu} are dimension independent; the decomposition
in equation (7) of the Riemann tensor in terms of the Weyl tensor,
Ricci tensor, and Ricci scalar is the first occurrence of
dimension-dependent coefficients. Note that in general $G_A{}^A =
-[(n-2)/2] R_A{}^A$.}
%------------------------------------------------------------------
% 
\begin{eqnarray}
{}^{(n-1)}G_{AB}=
&&\hspace{-6mm}
{n-3\over n-2}\left[
G_{CD}^\pm \; q_A{}^C \; q_B{}^D
+
\left(G_{CD}^\pm \; n^C n^D - {1\over n-1} G^\pm_C{}^C \right) \; q_{AB}
\right]
\nonumber \\
&&\hspace{-6mm}
+K^\pm \; K^\pm_{AB}-K^\pm_{A}{}^{C} K^\pm_{BC} - {1 \over 2}  
q_{AB}\; \left( K_\pm^2- K^\pm_{CD} K_\pm^{CD}\right) -E^\pm_{AB}.
\label{E:brane-einstein-tensor}
\end{eqnarray}
Here $E_{AB}$ is the ``electric'' part of the Weyl tensor
\begin{equation}
E_{AB} = 
C_{DEHI} \; n^D \; n^H \; q_A{}^E \; q_B{}^I =  
C_{BAHB} \; n^D \; n^H.
\end{equation}
Indeed, since the foliation we are dealing with is timelike (spacelike
normal) calling this the ``electric'' part of the Weyl tensor is
purely formal.

The above expression (\ref{E:brane-einstein-tensor}) is purely geometrical,
with as yet no physical content. Its meaning can be made clearer by
simultaneously substituting the bulk Einstein equations
\begin{equation}
G_{AB} = \kappa_n^{-2} \; T_{AB},
\end{equation}
and by adopting ``reduced'' Gaussian normal coordinates adapted to the
brane.  Using the definitions of the previous subsection equation
(\ref{E:brane-einstein-tensor}) becomes
\begin{eqnarray}
{}^{(n-1)}G_{ab}=
&&\hspace{-6mm}
{n-3\over n-2}\; \kappa_n^{-2} \; \left[
T_{ab}^\pm - {1\over n-1} (T_{cd}^\pm \; q^{cd}) \; q_{ab} 
+ {n-2\over n-1} \; T_{nn}^\pm
\right]
\nonumber \\
&&\hspace{-6mm}
+K^\pm \; K_{ab}^\pm-K^\pm_{a}{}^{c} K^\pm_{bc} - {1 \over 2}  
q_{ab}\; \left( K_\pm^2- K^\pm_{cd} K_\pm^{cd}\right) - C^\pm_{anbn}.
\label{E:brane-einstein-tensor-2}
\end{eqnarray}
You can take sums and differences of these two equations but the
results are not particularly enlightening---it is (finally) at this
stage that we find it useful to impose $Z_2$ symmetry (or adopt the
``one-sided'' approach of~\cite{void,edge} at the cost of introducing
a few factors of 2). In the case of $Z_2$ symmetry
\begin{equation}
K^+_{ab} = K_{ab} = - K^-_{ab}; \qquad \K_{ab} = 2 K_{ab} = 2 K^+_{ab} = -2 K^-_{ab}.
\end{equation}
For the ``one-sided'' approach~\cite{void,edge}
\begin{equation}
K^+_{ab} =  K_{ab}; \qquad 
K^-_{ab} = 0, \hbox{ ``null and void''}; \qquad 
\K_{ab} = K_{ab} = K^+_{ab}.
\end{equation}
In either case
\begin{eqnarray}
{}^{(n-1)}G_{ab}=
&&\hspace{-6mm}
{n-3\over n-2}\; \kappa_n^{-2} \; \left[
T_{ab} - {1\over n-1} (T_{cd} \; q^{cd}) \; q_{ab} 
+ {n-2\over n-1} \; T_{nn}
\right]
\nonumber \\
&&\hspace{-6mm}
+K \; K_{ab}-K_{a}{}^{c} \; K_{bc} 
- {1 \over 2}  
q_{ab}\; \left( K^2- K_{cd} K^{cd}\right) 
- C_{anbn}.
\label{E:brane-einstein-tensor-3}
\end{eqnarray}
In the original situation considered by
Shiromizu--Maeda--Sasaki~\cite{Shiromizu} the bulk stress energy was
particularly simple, $T_{AB} = -\Lambda \; g_{AB}$ so that $T_{ab} =
-\Lambda \; q_{ab}$ and $T_{nn} = -\Lambda$. In that case the terms on
the first line above are trivial (corresponding to a bulk-induced
contribution to the cosmological constant for the brane), while in the
second line the terms involving the extrinsic curvature were uniquely
determined by the brane stress-energy.  Thus in
Shiromizu--Maeda--Sasaki~\cite{Shiromizu} the {\emph{only}} way the
bulk enters into the effective equations of motion is via the Weyl
tensor contribution $C_{anbn}$.  In the recent extensions by Maeda and
Wands~\cite{Wands} and Mennim and Battye~\cite{Mennim} the bulk stress
tensor corresponds to a nontrivial dilaton field; and this opens up
new avenues for possible bulk influence on the brane.

In the more general case of this paper the bulk has in principle the
possibility of communicating with the brane in at least five different
ways: (1) directly via the Weyl tensor $C_{anbn}$, (2) via
``anisotropies'' in the in-brane components of the bulk stress-energy
[$T_{ab} - {1\over n-1} (T_{cd}\; q^{cd}) \; q_{ab}$], (3) via the
pressure $T_{nn}$, (4) via the now much more complicated relationship
between extrinsic curvature and surface stress-energy
(\ref{E:j-reassemble}), and (5) via the more complicated
``conservation law'' of the previous subsection
(\ref{E:non-conservation}).

By substituting the bulk Einstein tensor
(\ref{E:bulk-einstein-tensor}) in (\ref{E:brane-einstein-tensor-3}),
deferring for now the task of dealing with the extrinsic curvatures,
we obtain
\begin{eqnarray}
{}^{(n-1)}G_{AB}
&&\hspace{-6mm}
=
{n-3 \over n-2} \;F^{-1} \; 
\left[\kappa_n^{-2}\;H_{ij}+F''_{ij}\right] D_A \phi^i \; D_B \phi^j 
\nonumber \\
&&\hspace{-6mm}
- 
{n-3 \over n-2} \; F^{-1} \; q_{AB}
\left[{n\over 2(n-1) }\; \kappa_n^{-2}\; H_{ij}+ F''_{ij}\right] 
D_C \phi^i \; D^C \phi^j 
\nonumber \\
&&\hspace{-6mm}
+
{n-3 \over n-2} \; F^{-1} \;F'_i
\left[  D_A D_B \phi^i - q_{AB} \; 
D_C D^C \phi^i \right]  
\nonumber \\
&&\hspace{-6mm}
-
{n-3 \over n-1} \;(\Lambda + \kappa_n^{-2}\;F^{-1} \; V) \; q_{AB} 
\nonumber \\
&&\hspace{-6mm}
+ 
{n-3 \over n-1} \; \kappa_n^{-2}\; {1\over2} \; F^{-1} \;H_{ij} \;
(n^C\partial_C \phi^i) \; (n^D\partial_D \phi^j) \; q_{AB}
\nonumber \\
&&\hspace{-6mm}
-
{n-3 \over n-2} \; F^{-1} \;F'_i \; (n^A\partial_A \phi^i) \; (K_{AB}-K \; q_{AB})
\nonumber \\
&&\hspace{-6mm}
+
K\;K_{AB}-K_{A}{}^{C} \; K_{BC} 
- {1 \over 2} \; q_{AB}\;( K^2- K_{CD} K^{CD}) 
\nonumber \\
&&\hspace{-6mm}
-E_{AB}.
\label{E:brane-einstein-tensor-4}
\end{eqnarray}
The terms in the first four rows of (\ref{E:brane-einstein-tensor-4}) are
quantities intrinsically defined on the ($n$--1)-dimensional
hypersurface. ($D_A$ is the intrinsic covariant derivative on the
brane; acting on braneworld scalars $D_a \phi^i = q_A{}^B \nabla_B
\phi^i$; acting on tensors there are additional terms coming from the
extrinsic curvature.) The terms in the last four rows depend, a
priori, on the extrinsic properties of the hypersurface; we will now
use the junction conditions to rewrite them (as much as possible) in
terms of intrinsic quantities.

Recall that we are now restricting attention to the case of a $Z_2$
orbifold identification, (or, modulo a few factors of 2, adopting the
``one-sided'' approach of~\cite{void,edge}). The generalized junction
conditions (\ref{E:j-phi-2}) and (\ref{E:j-reassemble-2}) will then
read
\begin{equation}
(n^A\partial_A \phi^i)={1 \over 2}J^i=
{1\over 2 F} \left( 
\kappa_n^{-1} \; \gamma_{21}^i \; S +  \gamma_{22}^{ij} \; (\L_\brane)'_j
\right), 
\label{E:specialized-2}
\end{equation}
and
\begin{eqnarray}
K_{AB}
&=&{1 \over 2}\K_{AB}
%\\
%&=& 
=
{\kappa_n^{-2} \over 2 F } \; 
\left(
S_{AB} + \tilde\gamma_{11} \; S \; q_{AB} + 
\kappa_n \; \tilde\gamma_{12}^i \; (\L_\brane)'_i \; q_{AB}
\right).
\label{E:specialized-1}
\end{eqnarray}
These expressions can now be substituted into
(\ref{E:brane-einstein-tensor-4}) to give place to
($n$--1)-dimensional Einstein-like equations controlling the
gravitational behaviour inside the brane. We find that the most
efficient way to proceed is by introducing some generic coefficients
$\Gamma_1\dots \Gamma_7$. Doing this, and adopting ``reduced''
Gaussian normal coordinates
\begin{eqnarray}
{}^{(n-1)}G_{ab}=&&\hspace{-6mm}
{n-3 \over n-2} \;F^{-1} \; 
\left[\kappa_n^{-2}\;H_{ij}+F''_{ij}\right] D_a \phi^i \; D_b \phi^j 
\nonumber \\
&&\hspace{-6mm}
- 
{n-3 \over n-2} \; F^{-1} \; q_{ab}
\left[{n\over 2(n-1) }\; \kappa_n^{-2}\;H_{ij}+ F''_{ij}\right] 
D_c \phi^i \; D^c \phi^j 
\nonumber \\
&&\hspace{-6mm}
+{n-3 \over n-2} \; F^{-1} \;F'_i
\left[  D_a D_b \phi^i - q_{ab} \; 
D_c D^c \phi^i \right]  
\nonumber \\
&&\hspace{-6mm}
-{n-3 \over n-1}\;(\Lambda + \kappa_n^{-2}\;F^{-1} \; V) \; q_{ab} 
\nonumber \\
&&\hspace{-6mm}
+ {\kappa_n^{-4}\over4F^2} \left\{
\Gamma_1 \; (S^2)_{ab} 
+\Gamma_2 \; S \; S_{ab} 
+\Gamma_3 \; S^2 \; q_{ab} 
+\Gamma_4 \; (S_{cd}\;S^{cd}) \; q_{ab}
\right\}
\nonumber \\
&&\hspace{-6mm}
+ {\kappa_n^{-3}\over4F^2} \left\{ 
\Gamma_5^i  \; (\L_\brane)'_i \; S_{ab} 
+ \Gamma_6^i \; (\L_\brane)'_i \; S \; q_{ab} 
\right\}
\nonumber \\
&&\hspace{-6mm}
+ {\kappa_n^{-2}\over4F^2} \left\{ 
\Gamma_7^{ij}  \; (\L_\brane)'_i \;(\L_\brane)'_j   \; q_{ab}
\right\}
\nonumber \\
&&\hspace{-6mm}
-C_{anbn}.
\label{E:brane-einstein-tensor-5}
\end{eqnarray}
This serves as the definition of the dimensionless coefficients
$\Gamma_1 \dots \Gamma_7$; they are functions of the $\gamma$'s (and
thus implicitly functions of the on-brane values of the moduli fields)
and the dimensionality of spacetime.  Note that all the terms
appearing here are {\emph{intrinsic}} to the braneworld,
{\emph{except}} for the Weyl tensor $C_{anbn}$, so that, as explained
by Shiromizu--Maeda--Sasaki~\cite{Shiromizu}, this system of equations
is not closed. The Weyl term depends upon the global behaviour of the
bulk spacetime, which itself depends on the bulk scalar
field. 

Even without explicitly calculating the coefficients $\Gamma_1 \dots
\Gamma_7$ we see that several key features of the
Shiromizu--Maeda--Sasaki analysis carry through---such as the presence
of quadratic terms depending on the square of the braneworld
stress-energy tensor in these effective Einstein equations.  A tedious
but straightforward analysis leads to
\begin{eqnarray}
\Gamma_1 &=& - 1;
\\
\Gamma_2 &=& + {1\over n-2};
\\
\Gamma_3 &=&  -{1\over2(n-2)} + {n-3\over(n-1)^2(n-2)}\; {\E-1\over\E}
\\
&=& -{1\over2(n-2)}\left[1-  {n-3\over2(n-1)^2}\; {\E-1\over\E} \right];
\\
\Gamma_4 &=& + \half;
\\
\Gamma_5^i &=& 0;
\\
\Gamma_6^i &=& 
- {n-3\over(n-1)(n-2)} \; {\kappa_n \; H^{ij} \; F'_j \over \E} 
\\
&=& 
- {n-3 \over(n-1)(n-2)} \; {\kappa_n\; [H_E]^{ij} \; F'_j\over F};
\\
\Gamma_7^{ij} &=& +{1\over2} \;{n-3 \over n-1} \;[H_E^{-1}]^{ij};
\end{eqnarray}
where we have defined
\begin{eqnarray}
\E(\phi) &=& 
1  + 
{n-1\over n-2} \;H^{ij}(\phi) \;\kappa_n^2 \; 
{F'_i(\phi) \; F'_j(\phi) \over F(\phi)},
\end{eqnarray}
and $[H_E^{-1}]^{ij}$ is the inverse of the Einstein-frame sigma
model metric as defined in (\ref{E:sigma-model-metric}).

To facilitate comparison with the analyses of
Shiromizu--Maeda--Sasaki~\cite{Shiromizu}, Maeda--Wands~\cite{Wands},
and Mennim--Battye~\cite{Mennim} it is useful to split the surface
stress-energy into a (possibly moduli dependent) ``internal
cosmological constant'' ({\emph{not}} the ``effective cosmological
constant''---see below), and ``the rest'' according to the
prescription
\begin{equation}
S_{ab} = -\lambda(\phi) \; q_{ab} + \tau_{ab}(\phi).
\end{equation}
After this substitution, some rearrangement yields 
\begin{eqnarray}
{}^{(n-1)}G_{ab}=&&\hspace{-6mm}
{n-3 \over n-2} \;F^{-1} \; 
\left[\kappa_n^{-2}\;H_{ij}+F''_{ij}\right] D_a \phi^i \; D_b \phi^j 
\nonumber \\
&&\hspace{-6mm}
- 
{n-3 \over n-2} \; F^{-1} \; q_{ab}
\left[{n\over 2(n-1) }\; \kappa_n^{-2}\;H_{ij}+ F''_{ij}\right] 
D_c \phi^i \; D^c \phi^j 
\nonumber \\
&&\hspace{-6mm}
+{n-3 \over n-2} \; F^{-1} \;F'_i
\left[  D_a D_b \phi^i - q_{ab} \; 
D_c D^c \phi^i \right]  
\nonumber \\
&&\hspace{-6mm}
-{n-3 \over n-1} \;(\Lambda + \kappa_n^{-2}\;F^{-1} \; V) \; q_{ab} 
\nonumber \\
&&\hspace{-6mm}
+ {\kappa_n^{-4}\over4F^2} \left\{
\Gamma_1 \; (\tau^2)_{ab} 
+\Gamma_2 \; \tau \; \tau_{ab} 
+\Gamma_3 \; \tau^2 \; q_{ab} 
+\Gamma_4 \; (\tau_{cd}\;\tau^{cd}) \; q_{ab}
\right\}
\nonumber \\
&&\hspace{-6mm}
+ {\kappa_n^{-4}\over4F^2} \left\{
-2\; \Gamma_1 \; \lambda 
-(n-1) \;\Gamma_2 \; \lambda
+\kappa_n \; \Gamma_5^i  \; (\L_\brane)'_i  
\right\} \; \tau_{ab}
\nonumber \\
&&\hspace{-6mm}
+ {\kappa_n^{-4}\over4F^2} \left\{
-\Gamma_2  \; \lambda 
-2\;(n-1)\;\Gamma_3 \; \lambda
-2\;\Gamma_4 \; \lambda 
+\kappa_n \; \Gamma_6^i  \; (\L_\brane)'_i  
\right\} \; \tau \; q_{ab}
\nonumber \\
&&\hspace{-6mm}
+ {\kappa_n^{-4}\over4F^2} 
\Big\{
\Gamma_1  \; \lambda^2 
+ (n-1) \; \Gamma_2 \; \lambda^2
+ (n-1)^2 \;\Gamma_3 \; \lambda^2
+ (n-1) \;  \Gamma_4 \; \lambda^2 
\nonumber\\
&&
\qquad
-\kappa_n \; \Gamma_5^i  (\L_\brane)'_i  \; \lambda
-\kappa_n \; \Gamma_6^i  (\L_\brane)'_i  \; (n-1) \; \lambda
\nonumber\\
&&
\qquad
+\kappa_n^2 \; \Gamma_7^{ij} \; (\L_\brane)'_i \;  (\L_\brane)'_j
\Big\} \; q_{ab}
\nonumber \\
&&\hspace{-6mm}
-C_{anbn}.
\label{E:brane-einstein-tensor-6}
\end{eqnarray}
To interpret this physically, we write it as
\begin{eqnarray}
{}^{(n-1)}G_{ab} &=& 
8\pi\;G_\effective \; \tau_{ab} 
+ 
8\pi\;G_\anomalous \; \tau \; q_{ab} 
-
\Lambda_\effective \; q_{ab}
\nonumber\\
&&
+ 8\pi\;G_\quadratic \;  
\left\{
\Gamma_1 \; (\tau^2)_{ab} 
+
\Gamma_2 \; \tau \; \tau_{ab} 
+
\Gamma_3 \; \tau^2 \; q_{ab} 
+
\Gamma_4 \; (\tau_{cd}\;\tau^{cd}) \; q_{ab}
\right\}
\nonumber\\
&&
+\kappa_n^{-2} \; (T^\phi)_{ab} - C_{anbn}.
\label{E:physical}
\end{eqnarray}
Here $G_\effective$ denotes the effective Newton constant on the
brane, in general it is a function of the moduli
fields. $G_\anomalous$ represents a perhaps unexpected anomalous
coupling of the braneworld geometry to the trace of braneworld stress
energy: in semi-realistic models this should be made small (and in
standard scenarios it often vanishes identically; more on this
later). $\Lambda_\effective$ is the net effective cosmological
constant for matter trapped on the brane---it gets contributions from
both the bulk and the brane, according to the formula given below; in
realistic models it should be kept as small as
possible. $G_\quadratic$ governs the quadratic contributions to the
effective on-brane Einstein equations; again in realistic models this
quantity should be kept small to avoid serious conflict with
experiment and observational cosmology. Finally $(T^\phi)_{ab}$
represents the effective stress-energy attributable to along-the-brane
variations of the moduli fields. Explicitly
\begin{eqnarray}
8\pi\;G_\effective &=&  
{\kappa_n^{-4}\over4F^2} 
\left\{
-2 \; \Gamma_1 \; \lambda 
-(n-1) \;\Gamma_2 \;  \lambda
+\kappa_n \; \Gamma_5^i  \; (\L_\brane)'_i  
\right\};
\\
8\pi\;G_\anomalous &=&  
{\kappa_n^{-4}\over4F^2} 
\left\{
-\Gamma_2  \; \lambda 
-
2 \; (n-1) \;\Gamma_3 \; \lambda
-
2\; \Gamma_4 \; \lambda 
+\kappa_n \; \Gamma_6^i \; (\L_\brane)'_i  
\right\};
\\
8\pi\;G_\quadratic &=& 
{\kappa_n^{-4}\over4F^2}; 
\\
\Lambda_\effective &=&
{n-3 \over n-1} \;(\Lambda + \kappa_n^{-2}\;F^{-1} \; V)
\nonumber\\
&&
- {\kappa_n^{-4}\over4F^2} 
\Big\{
\Gamma_1  \; \lambda^2 
+ (n-1)    \;\Gamma_2 \;\lambda^2
+ (n-1)^2  \;\Gamma_3 \;\lambda^2
+ (n-1)    \;\Gamma_4 \;\lambda^2 
\nonumber\\
&&
\quad
-\kappa_n \; \Gamma_5^i  \;(\L_\brane)'_i  \; \lambda
-\kappa_n \; (n-1) \; \Gamma_6^i  \;(\L_\brane)'_i  \; \lambda
\nonumber\\
&&
\quad
+\kappa_n^2 \; \Gamma_7^{ij}  \;(\L_\brane)'_i   \;(\L_\brane)'_j
\Big\}.
\end{eqnarray}
We draw some general largely model-independent conclusions:
Generically, the induced Einstein equations on the braneworld
correspond to a generalization of the notion of a Brans--Dicke theory,
with an effective Newton constant that depends on possibly
position-dependent scalar fields (the moduli) which themselves
influence both the effective cosmological constant and make direct
contributions to the stress-energy. Where the braneworld approach
steps well beyond the usual Brans--Dicke theories is in situations
where there is significant coupling to the bulk---either through
incoming fluxes or a nontrivial bulk Weyl tensor.

Inserting the explicit formulae for $\Gamma_1\dots\Gamma_7$ we find
\begin{eqnarray}
8\pi\;G_\effective &=&  
+ {\kappa_n^{-4}\; \lambda \over4 F^2} \; {n-3\over n-2};
\\
8\pi\;G_\anomalous &=&  
- {\kappa_n^{-4}\over4 F^2} \; {n-3\over(n-1)(n-2)}
\left\{
\lambda\;{\E-1\over\E} + {\kappa_n^2 \;F'_i \;H^{ij} \;(\L_\brane)'_j \over \E}
\right\}
\\
&=&
- {\kappa_n^{-4}\over4 F^2} \; {n-3\over(n-1)(n-2)}
\left\{
\lambda\;{\E-1\over\E} + {\kappa_n^2 \;F'_i \;[H_E^{-1}]^{ij} \;(\L_\brane)'_j \over F}
\right\};
\\
8\pi\;G_\quadratic &=& 
+ {\kappa_n^{-4}\over4 F^2}; 
\\
\Lambda_\effective &=&
{n-3 \over n-1} \;(\Lambda  + \kappa_n^{-2} \; F^{-1} \; V)
%\nonumber
%\\
%&&
+ {\kappa_n^{-4} \; \lambda^2 \over8 F^2} \; {n-3\over n-2} \; {1\over\E}
\nonumber
\\
&&
-{\kappa_n^2\over 4 F^2} \; {n-3\over n-2} {[H_E]^{ij}  F'_i   \;(\L_\brane)'_j\over F}
\nonumber
\\
&&
-{\kappa_n^2\over8 F^2} \; {n-3\over n-1} [H_E]^{ij}  \;(\L_\brane)'_i   \;(\L_\brane)'_j
\Big\}.
\end{eqnarray}
The ``quadratic'' part of the effective stress tensor is proportional
to
\begin{equation}
- (\tau^2)_{ab} 
+ {\tau \; \tau_{ab}\over n-2} 
- {\tau^2 \; q_{ab}\over2(n-2)} \left[ 1 - {n-3\over (n-1)^2} {\E-1\over\E} \right] 
+ \half (\tau^{pq} \; \tau_{pq}) \; q_{ab}.
\end{equation}
Finally the explicit moduli contribution to the effective stress
tensor appearing in equation (\ref{E:physical}) is
\begin{eqnarray}
(T^\phi) _{ab} &=& 
{n-3 \over n-2} \;F^{-1} \; 
\left[H_{ij}+\kappa_n^{2}\;F''_{ij}\right] D_a \phi^i \; D_b \phi^j 
\nonumber \\
&&\hspace{-6mm}
- 
{n-3 \over n-2} \; F^{-1} \; q_{ab}
\left[{n\over 2(n-1) }\; H_{ij}+ \kappa_n^{2}\;F''_{ij}\right] 
D_c \phi^i \; D^c \phi^j 
\nonumber \\
&&\hspace{-6mm}
+{n-3 \over n-2} \; \kappa_n^{2} \; F^{-1} \;F'_i
\left[  D_a D_b \phi^i - q_{ab} \; 
D_c D^c \phi^i \right].
\label{E:t-moduli} 
\end{eqnarray}

Some key points to realize are: 
\\
---(1) As long as $\lambda$ is positive (corresponding to a positive
brane tension), the effective Newton constant $G_\effective$ is also
positive, with the sign being independent of total
dimensionality. While the use of negative tension branes in supporting
roles has gained considerable popularity (see references
in~\cite{void}), the use of negative tension branes should be viewed
with extreme suspicion: Not only is the effective braneworld Newton
constant negative for matter trapped on the brane, but negative
tension branes also cause disturbing effects in the bulk. It has been
known for over a decade that negative tension branes led to
traversable wormholes~\cite{surgery}, with all their attendant
problems~\cite{book}, a point that we have made more explicit in a
braneworld context in~\cite{void}.
\\
---(2) $G_\anomalous$ vanishes whenever $F'_i(\phi)=0$; in particular,
it automatically vanishes in the Einstein frame. This is the reason
this contribution has been invisible to date.
\\
---(3) Quadratic terms are unavoidable. They cannot be removed by
change of conformal frame or choice of dimensionality. The best that
can be done is to set the coefficient of one of the quadratic terms
($\tau^2)$ to zero by fine tuning $\E=-(n-3)/(n^2-3n+4)$.
\\ 
---(4) $\Lambda_\effective$ is now much more complicated. The effective
cosmological constant can be modified by tuning the moduli fields.

In summary: Many of the qualitative features of the
Shiromizu--Maeda--Sasaki analysis~\cite{Shiromizu}, plus the
extensions by Maeda--Wands~\cite{Wands}, and
Mennim--Battye~\cite{Mennim}, continue to hold in this much more
general framework.  To illustrate what we can learn from the induced
braneworld Einstein equations let us now consider some specific
examples: (1) we particularize to the Einstein frame (keeping
dimensionality, sigma-model, number of moduli fields, and the brane
Lagrangian generic); as should be expected, we encounter considerable
simplifications, (2) we consider the dilaton field in the string
frame, again keeping the discussion as general as is reasonable.

%-----------------------------------------------------------------------
\section{Generic moduli fields in the Einstein frame}
%-----------------------------------------------------------------------
\setcounter{equation}{0}
%---------------------------------------------------------------------

Suppose we decide to do all calculations in the Einstein frame, but
keep the moduli fields arbitrary in all other respects. We have already
seen that going to the Einstein frame casts the general action into the
considerably simpler form (\ref{E:action-in-einstein-frame}):
\begin{eqnarray}
\S_\Einstein=&&\hspace{-6mm}
{1 \over 2}\int_{{\rm int}(\M)} \sqrt{-g_E} \;\d^n x \; \kappa_n^{2} \; 
\left[ R(g_E) -2 F(\phi)^{-2/(n-2)} \; \Lambda \right] 
\nonumber\\
&&
-\int_{\partial\M} \sqrt{-q_E} \;\d^{n-1}x \;  \kappa_n^{2} \;K(g_E)
\nonumber\\  
&&\hspace{-6mm}
+\int_{{\rm int}(\M)} \sqrt{-g_E} \;\d^n x \;
\Bigg\{ -{1 \over 2} 
[H_E]_{ij}(\phi) \; 
\left[ g^{AB} \; \partial_{A}\phi^i \; \partial_{B}\phi^j \right] 
\nonumber
\\
&&
\qquad
- V_E(\phi,\psi) 
\nonumber
\\
&&
\qquad
+ \L_\bulk^E([g_E]_{AB},\phi,\psi)
\Bigg\}
\nonumber \\  
&&\hspace{-6mm}
+\int_{\rm brane} \sqrt{-q_E} \;\d^{n-1}x\; 
\L_\brane^E([q_E]_{AB},\phi,\psi).
\label{E:action-in-einstein-frame-simplified}
\end{eqnarray} 
Various factors of $F(\phi)$ are now hiding in the definition of
$V_E$, $\L_\bulk$, and $\L_\brane$.  We explicitly see that what was
the cosmological constant in the original frame is now a potential.
Similarly if we had a simple brane tension in the original frame, then
in this new frame the brane tension will be moduli dependent.
(Warning: Because of this you cannot just blindly set $F(\phi)=1$
everywhere and hope to get meaningful results, going to the Einstein
frame is a field redefinition which simplifies the gravitational sector
but there is a ``conservation of difficulty'' phenomenon and the bulk
and brane matter Lagrangians are more complicated). The bulk
stress-energy tensor simplifies to
\begin{eqnarray}
[T_E]_{AB}^\effective &=& 
[H_E]_{ij}(\phi) \; \partial_{A}\phi^i \; \partial_{B}\phi^j 
\nonumber \\
&& - 
{1 \over 2} \;
[H_E]_{ij}(\phi) \; [g_E]_{AB} \;\;
[g_E]^{CD} \;\partial_{C}\phi^i \; \partial_{D}\phi^j 
\nonumber \\
&&
+ [T_E^\psi]_{AB}
- [g_E]_{AB} V_E(\phi,\psi)
- [g_E]_{AB}\; \kappa_n^2 \; \Lambda  \; F(\phi)^{-2/(n-2)} .
\label{E:bulk-einstein-frame}
\end{eqnarray}
Note that this stress-energy tensor is defined by variation with
respect to $g_E$ and does not necessarily have any simple relationship
to the original stress-energy tensor $T_{AB}$; the same comment
applies to the surface stress tensor $[S_E]_{AB}$.  Once this
redefinition is performed, and provided we agree to phrase questions
in terms of $[T_E]_{AB}$ and $[S_E]_{AB}$, the junction conditions
also simplify (since there is no longer any mixing between $J^i$ and
$K$). We get
\begin{equation}
\K_E= -{\kappa_n^{-2} \; S_E\over n-2},
\end{equation}
and
\begin{eqnarray} 
J^i &=& - [H_E]^{ij} \; (\L_\brane^E)'_j.
\end{eqnarray} 
As previously noted, the Codazzi equation simplifies to
(\ref{E:codazzi-in-einstein}). Finally the Gauss equation implies that
the braneworld geometry satisfies
\begin{eqnarray}
{}^{(n-1)}[G_E]_{AB}
&&\hspace{-6mm}
=
{n-3 \over n-2} \;\kappa_n^{-2}\;[H_E]_{ij}\; D_A \phi^i \; D_B \phi^j 
\nonumber \\
&&\hspace{-6mm}
- 
{n(n-3) \over 2(n-1)(n-2)} \; [q_E]_{AB} \; \kappa_n^{-2}\;\; 
[H_E]_{ij} \;D_C \phi^i \; D^C \phi^j 
\nonumber \\
&&\hspace{-6mm}
-
{n-3 \over n-1} \;(\Lambda\;  F(\phi)^{-2/(n-2)} + \kappa_n^{-2} \; V_E) \; [q_E]_{AB} 
\nonumber \\
&&\hspace{-6mm}
+ 
{n-3 \over n-1} \; \kappa_n^{-2}\; {1\over2} \;[H_E]_{ij} \;
(n^C\partial_C \phi^i) \; (n^D\partial_D \phi^j) \; [q_E]_{AB}
\nonumber \\
&&\hspace{-6mm}
+
K\;K_{AB}-K_{A}{}^{C} \; K_{BC} 
- {1 \over 2} \; [q_E]_{AB}\;( K^2- K_{CD} K^{CD}) 
\nonumber \\
&&\hspace{-6mm}
-E_{AB}.
\label{E:brane-einstein-einstein-frame}
\end{eqnarray}
(With the extrinsic curvature and the Weyl tensor being calculated
using the Einstein frame metric.)  As in the general case, you can
introduce dimensionless coefficients $\Gamma_1\dots\Gamma_7$,
rearrange terms, and introduce an effective Newton constant and
effective brane cosmological constant. These are now relatively simple
functions of the dimensionality and of the background moduli fields.
We collect some technical results in Appendix B, and here merely quote
the results for the effective Newton constant and related parameters:
\begin{eqnarray}
8\pi\;G_\effective &=&  
{\kappa_n^{-4}\; \lambda\over4} {n-3\over n-2};
\\
8\pi\;G_\anomalous &=&  0:
\\
8\pi\;G_\quadratic &=& 
{\kappa_n^{-4}\over4}; 
\\
\Lambda_\effective &=&
{n-3 \over n-1} \;(\Lambda \; F^{-2/(n-2)} + \kappa_n^{-2} \; V_E)
\nonumber\\
&&
+ {\kappa_n^{-4}\over8} 
\Big\{ 
{n-3\over n-2} \lambda^2 
-\kappa_n^2 \; {n-3\over n-1} [H_E]^{ij}  \;(\L_\brane)'_i   \;(\L_\brane)'_j
\Big\}.
\end{eqnarray}
In particular note that $G_\anomalous=0$.  Furthermore the effective
cosmological constant picks up contributions from whatever moduli
dependence the brane Lagrangian may posses.  The coefficients of the
``quadratic'' pieces of induced Einstein equations are simply
\begin{equation}
- (\tau^2)_{ab} 
+ {\tau \; \tau_{ab}\over n-2} 
- {\tau^2 \; q_{ab}\over2(n-2)}
+ \half (\tau^{pq} \; \tau_{pq}) \; q_{ab},
\end{equation}
while the explicit moduli contribution to the effective stress
tensor appearing in equation (\ref{E:physical}) now reduces to
\begin{eqnarray}
(T^\phi) _{ab} &=& 
{n-3 \over n-2} \; 
[H_E]{ij}\; D_a \phi^i \; D_b \phi^j 
- 
{n(n-3) \over2 (n-2)(n-1)} \; q_{ab} \; [H_E]{ij} \; 
D_c \phi^i \; D^c \phi^j. 
\label{E:t-moduli-einstein} 
\end{eqnarray}

In comparing with the Maeda--Wands analysis~\cite{Wands} note there
are several subtle differences in normalization and sign convention
(we have tried to stick to MTW conventions~\cite{MTW}) and that they
have specifically chosen the in-brane cosmological constant to be the
only moduli-dependent piece of the brane Lagrangian, so that they have
\begin{equation}
\L_\brane([g_e],\phi,\psi) = - \lambda(\phi) + \L_\brane([g_E],\psi).
\end{equation}
Consequently in their analysis 
\begin{equation}
(\L_\brane)'_i  = - \lambda'(\phi) = - {\d\lambda\over\d\phi}.
\end{equation}
Modulo choices of convention, our results (when specialized to $n=5$,
and a single dilaton field with canonical kinetic energy) are in
agreement with theirs. See equations (2.18)--(2.22) of~\cite{Wands}

Similarly in comparing with the Mennim--Battye analysis~\cite{Wands}
note there are other subtle differences in normalization and sign
convention, in particular their $V \to -2\Lambda$ and their $U \to
-2\lambda(\phi)$ in our notation. Then consider for instance equation
(22) of~\cite{Mennim} and compare with our more general formula for the
effective cosmological constant as presented above.

%-----------------------------------------------------------------------
\section{Dilaton field in the string frame}
%-----------------------------------------------------------------------
\setcounter{equation}{0}
%---------------------------------------------------------------------

Consider $F(\phi)=\exp(-2\phi/\kappa_n)$, with
$H(\phi)=-4\;\exp(-2\phi/\kappa_n)$. This corresponds to the dilaton
field in the string frame~\cite{generalized-junction}. Some key
coefficients are collected in Appendix C.  The generalized junction
conditions become
\begin{eqnarray}
\K_{ab}&=& \exp(2\phi/\kappa_n)  \; \kappa_n^{-2}
\left\{
S_{ab}+{\kappa_n \over 2} \; (\L_\brane)' \; q_{ab}
\right\},
\\
J&=& \exp(2\phi/\kappa_n)  \; \kappa_n^{-1}
\left\{
{1 \over 2} S -{n-2 \over 4}\; \kappa_n \; (\L_\brane)'b.
\right\}.
\end{eqnarray}
Then in particular
\begin{equation}
\K_{ab} - \K \; q_{ab} = \exp(2\phi/\kappa_n) 
\left\{
S_{ab} - S\; q_{ab} - {n-2\over2} \kappa \; (\L_\brane)' \; q_{ab}
\right\}.
\end{equation}
From the Codazzi equation the ``conservation'' of braneworld
stress-energy reads
\begin{equation}
D^a \left\{  \exp(2\phi/\kappa_n) \left[
S_{ab} - S\; q_{ab} - {n-2\over2} \; \kappa_n \; (\L_\brane)' \; q_{ab}
\right]
\right\} = - \F_b.
\end{equation}
So if you define the braneworld stress-energy by variation with
respect to the string metric, that particular stress-tensor is
{\emph{not}} conserved, both due to explicit interchange of stress
energy with the bulk, and (ultimately due to the nontrivial
matter-dilaton couplings in the string frame) due to possible
variations of the dilaton field along the brane.  Even if the dilaton
field is constant along the brane (with both $D_a \phi=0$, and
$D_a[(\L_\brane)'_i]=0$, appropriate for a ``translationally
invariant'' ground state), one still gets the perhaps unexpected
result
\begin{equation}
D^a \left\{ 
S_{ab} - S\; q_{ab}
\right\} = - \F_b.
\end{equation}
It is only if the braneworld stress-energy is additionally traceless,
and if there is no net flux onto the brane, that one recovers the
naive result
\begin{equation}
D^a \left\{ 
S_{ab}
\right\} = 0.
\end{equation}
If you are trying to do braneworld cosmology in the string frame, you
obtain results you might naively expect during the radiation dominated
expansion of the universe, but would see what appears to be
stress-energy nonconservation during the matter dominated era. This is
not an ``error'' or an ``inconsistency'' but merely a reflection of
the fact that (after taking account of reduction from the bulk to the
brane and coupling to the dilation) the string frame braneworld
stress-energy tensor does not quite have the properties you might
naively expect.

It is now straightforward (given the formalism developed herein) to
calculate the braneworld Einstein tensor (\ref{E:physical}).  For the
effective Newton constant and related quantities
\begin{eqnarray}
8\pi\;G_\effective &=&  
{\kappa_n^{-4}\; \exp(4\phi/\kappa_n) \;\lambda\over4} 
\left\{
{n-3\over n-2}
\right\};
\\
8\pi\;G_\anomalous &=&   
{\kappa_n^{-4}\; \exp(4\phi/\kappa_n)\over4} 
\Big\{ 
-\lambda {n-3\over n-2} + {n-3\over2(n-1)}\; \kappa_n \;(\L_\brane)'
\Big\}; 
\\
8\pi\;G_\quadratic &=& 
{\kappa_n^{-4}\; \exp(4\phi/\kappa_n) \over4}; 
\\
\Lambda_\effective &=&
{n-3 \over n-1} \;
\left[ \Lambda  + \kappa_n^{-2} \exp(2\phi/\kappa_n)\; V(\phi) \right]
\\
&-& 
(n-3)
{\kappa_n^{-4}\; \exp(4\phi/\kappa_n)\over8} 
\Big\{  
\lambda^2 
+\lambda\; \kappa_n \;(\L_\brane)'
+ {n-2\over4(n-1)}   \kappa_n^2 \;\;[(\L_\brane)']^2
\Big\}.
\nonumber
\end{eqnarray}
The ``quadratic'' part of the effective stress tensor is different from
that occurring in the Einstein frame and is now proportional to
\begin{equation}
- (\tau^2)_{ab} 
+ {\tau \; \tau_{ab}\over n-2} 
- {\tau^2 \; q_{ab}\over(n-1)(n-2)} 
+ \half (\tau^{pq} \; \tau_{pq}) \; q_{ab}.
\end{equation}
Finally the explicit moduli contribution to the effective stress
tensor appearing in equation (\ref{E:physical}) is
\begin{eqnarray}
(T^\phi) _{ab} &=& 
- 2 {n-3 \over n-1} \; q_{ab} \; D_c \phi^i \; D^c \phi^j 
-2{n-3 \over n-2} \; \kappa_n
\left[  D_a D_b \phi^i - q_{ab} \; D_c D^c \phi^i \right].
\label{E:t-moduli-string} 
\end{eqnarray}

Some points to note:
\\
---(1) You still need positive brane tension to make effective gravity
in the braneworld attract.
\\
---(2) $G_\anomalous$ is generally nonzero though it can be fine-tuned
away if you enforce
\begin{eqnarray}
(\L_\brane)' = 2 \;{n-1\over n-2} \; \lambda.
\end{eqnarray}
---(3) Predicting even the {\emph{sign}} of the brane contribution to
$\Lambda_\effective$ is now a lot trickier.

Let us now consider a more specific example: A ``bare'' brane (no
extra matter apart from the brane tension) with a Lagrangian
$\L_\brane=- \lambda(\phi)$, and thus with $S_{ab}=-\lambda(\phi) \;
q_{ab}$.  The Einstein equation for the brane becomes
\begin{equation}
{}^{(4)}G_{ab}=-\Lambda_\effective \; q_{ab},
\end{equation}
with 
\begin{equation}
\Lambda_{\effective}=
{n-3 \over n-1} \Lambda -
{(n-3)\kappa_n^{-4}\over8} \; \exp(4\phi/\kappa_n) \;
\left\{
\lambda(\phi)^2+ 
\lambda(\phi)\;\kappa_n\;\lambda'(\phi)+ 
{n-2\over 4(n-1)} \kappa_n^2 [\lambda'(\phi)]^2   
\right\}.
\end{equation}
Making an ansatz for the coupling interaction function of the form
$\lambda(\phi) = \lambda_0\;\exp(-\alpha\phi/\kappa_n)$, we can see
that 
\begin{equation}
\Lambda_{\effective}=
{n-3 \over n-1} \Lambda -
{(n-3)\kappa_n^{-4}\over8} \; \exp(4\phi/\kappa_n) \; \lambda_0^2
\left\{
1 - \alpha + {n-2\over 4(n-1)} \alpha^2  
\right\}.
\end{equation}
Then for $\alpha$ between $2[(n-1)\pm\sqrt{n-1}]/(n-2)$ there is a
positive contribution to the effective cosmological constant coming
from the brane tension. For other values of $\alpha$ there is a
negative contribution to the effective cosmological constant. In order
to find a solution in which there is a Poincare invariant brane
($\Lambda_{\effective}=0$) when $\alpha\in
2[(n-1)\pm\sqrt{n-1}]/(n-2)$ the bulk cosmological constant must be
negative (this is the usual situation, corresponding to an
anti-de~Sitter bulk).  That is the case, for example, when
$\alpha=2(n-1)/(n-2)$ (which also corresponds to fine-tuning
$G_\anomalous$ to zero). On the other hand, when for example
$\alpha=0$, one would need a positive bulk cosmological constant (a
de~Sitter bulk) if one wishes to accommodate a Poincare invariant
brane.

[With hindsight this should not be all that surprising, and in fact
the same phenomenon (with slightly different coefficients) also shows
up in the Einstein frame. Specifically pick canonical kinetic energies
$H_E=1$ and keep the brane tension ansatz $\lambda(\phi) =
\lambda_0\;\exp(-\alpha\phi/\kappa_n)$ used here. Then for $\alpha$ between
$\pm\sqrt{(n-1)/(n-2)}$ you need an anti-de Sitter bulk, while for
$\alpha$ outside this region you would need a de Sitter bulk to obtain
a Poincare invariant brane.]

%-----------------------------------------------------------------------
\section{Discussion}
%-----------------------------------------------------------------------
\label{S:discussion}
%-----------------------------------------------------------------------
\setcounter{equation}{0}
%---------------------------------------------------------------------

In this paper we have developed a general technique for analyzing
braneworld geometry in the presence of an arbitrary number of bulk
moduli fields. We also permit the use of arbitrary conformal frames,
since sometimes one conformal frame may be more useful than others. If
one is interested primarily (or solely) in gravitational phenomena,
the Einstein frame is often the best choice. For particle physics in
Brans--Dicke theories (and their generalizations) the Jordan frame is
often the best choice. String theory does not posses a unique Jordan
frame, but the string frame is perhaps the best analog in stringy
models.

We find that while many of the results known from situations where the
dilaton is frozen out by hand, or when one {\emph{ab initio}}
restricts attention to an Einstein frame formulation, continue to
hold in this more general context. On the other hand, several things
change: (1) in general frames there is a possibility of an anomalous
coupling between the trace of stress-energy and braneworld gravity;
(2) there is the potential for the exchange of stress-energy and
information between the bulk, the on-brane variations of the moduli
fields, and the braneworld stress-energy; (3) for moduli-dependent
brane tensions, the relationship to the effective cosmological
constant is more complex, and in particular one can flip the sign of
the brane-tension contribution to the effective Newton constant.

While we have analyzed the behaviour of the gravity sector in some
detail, it should be emphasised that there is a lot of flexibility in
the class of models we consider. For instance, the use of a generic
sigma model for the moduli fields gives you a lot of freedom. Likewise
the bulk potential $V(\phi)$ and moduli-dependent brane tension
$\lambda(\phi)$ are freely specifiable in this formalism. (More
generally, $\L_\brane(g,\phi,\psi)$ and $\L_\bulk(g,\phi,\psi)$ are
freely specifiable---apart from the fact that they should not contain
derivatives of $\phi$, because if so they would generate additional
terms in the ``momenta'' used to set up the generalized junction
conditions.) We have also had essentially nothing specific to say
about the ``matter'' fields that are assumed to be trapped on or near
the brane; from the current perspective they are simply a minor
contaminant to be dealt with after the large scale braneworld geometry
has been deduced from the interaction between brane tension, bulk
moduli fields, and bulk gravity.

Finally, in order to facilitate the analysis of a braneworld cascade,
branes within branes within branes, we have kept the dimensionality
of the bulk arbitrary.

%------------------------------------------------------------------------------
\section*{Acknowledgments}
%-----------------------------------------------------------------------------
The research of CB was supported by the Spanish Ministry of Education
and Culture (MEC). MV was supported by the US Department of Energy.

%-----------------------------------------------------------------------
\appendix
%-----------------------------------------------------------------------
\section*{Appendix A: Some coefficients (general frame)}
%-----------------------------------------------------------------------
\label{S:coefficents}
%-----------------------------------------------------------------------
\setcounter{equation}{0}
%---------------------------------------------------------------------
\renewcommand{\theequation}{A.\arabic{equation}}
%-----------------------------------------------------------------------

As we have seen in the general formula for $\Gamma_1\dots\Gamma_7$, it
is useful to define
\begin{eqnarray}
\E(\phi) &=& 
1  + 
{n-1\over n-2} \;H^{kl}(\phi) \;\kappa_n^2 \; 
{F'_k(\phi) \; F'_l(\phi) \over F(\phi)}.
\end{eqnarray}
This coefficient shows up repeatedly---ultimately this is due to the
fact that $\E$ is intimately related to the determinant of the sigma
model metric in the Einstein frame ($\#$ denotes the total number of moduli fields)
\begin{equation}
\det[H_E]  = \det[H] \; F^{-\#} \; \E.
\end{equation}
As such its vanishing is key to the ``exceptional'' case of ``Cheshire
cat branes'' (phantom branes) considered
in~\cite{generalized-junction}.  It also occurs as a sub-piece of the
various coefficients enumerated below:
\begin{eqnarray}
\gamma_{11} &=& 
- {1\over n-2} \; \E^{-1};
\\
\gamma_{12}^i &=& 
- {n-1\over n-2} \; H^{ij} \; \; \kappa_n F'_j \; \E^{-1};
\\
\gamma_{21}^i &=&  
+  {1\over n-2} \; H^{ij} \; \; \kappa_n F'_j \; \E^{-1};
\\
\gamma_{22}^{ij} &=& 
-  
F \left( 
H^{ij} - 
{n-1\over n-2} 
{\kappa_n^2 \; (H^{ik} \; F'_k) \; (H^{jl} \; F'_l) \over F \; \E}
\right) 
= - (H_E^{-1})^{ij}.
\end{eqnarray}
(The explicit factors of $\kappa_n$ keep all these coefficients
dimensionless.)  Some derived quantities are
\begin{eqnarray}
\tilde\gamma_{11} &=& 
{\gamma_{11}-1\over n-1} 
\\
&=& - {1\over n-2} +{\E-1\over\E(n-1)(n-2)}
\\
&=& - {1\over n-2}\; \left[1- {\E-1\over\E(n-1)}\right]
\\
&=& - {1\over\E(n-2)} -{\E-1\over\E(n-1)};
\\
\tilde\gamma_{12}^i &=& {1\over n-1} \gamma_{12}^i 
= - \gamma_{21}^i 
\\
&=& - {1\over n-2} \; H^{ij} \; \; \kappa_n F'_j \; \E^{-1} 
\\
&=& - {1\over n-2} \; [H_E^{-1}]^{ij} \; \; \kappa_n F'_j \; F^{-1} ;
\\
\alpha &=&  -{\kappa_n \; F'_i\over F} \gamma^i_{21} 
= -  {1\over n-2} \; F^{-1} 
\;H^{ij} \; \kappa_n^2  \;F'_i  \;F'_j \; \E^{-1} = -{\E-1\over\E(n-1)};
\\
\beta^i &=& - {\kappa_n \; F'_j\over F} \gamma_{22}^{ij} 
=  +  [H_E^{-1}]^{ij} \; \; \kappa_n F'_j \; F^{-1}
=  + H^{ij} \; {\kappa_n \; F'_j} \; \E^{-1}.
\end{eqnarray}
The $\Gamma_1\dots\Gamma_7$ coefficients given in the main body of the
paper were derived by substitution and rearrangement of
(\ref{E:specialized-1}}) and (\ref{E:specialized-2}) into
(\ref{E:brane-einstein-tensor-4}) using these $\gamma$
coefficients. The calculation in a general frame is tedious, though in
the Einstein frame it is relatively simple.

%-----------------------------------------------------------------------
\appendix
%-----------------------------------------------------------------------
\section*{Appendix B: Coefficients in the Einstein frame}
%-----------------------------------------------------------------------
\label{A:coefficents-in-einstein}
%-----------------------------------------------------------------------
\setcounter{equation}{0}
%---------------------------------------------------------------------
\renewcommand{\theequation}{B.\arabic{equation}}
%-----------------------------------------------------------------------

Einstein frame calculation:
\begin{eqnarray}
\E &=& 1;
\\
\gamma_{11} &=& - {1\over n-2};
\\
\gamma_{12}^i &=& 0;
\\
\gamma_{21}^i &=& 0;
\\
\gamma_{22}^{ij} &=&  - (H_E^{-1})^{ij}.
\\
\tilde\gamma_{11} &=& {\gamma_{11}-1\over n-1} = - {1\over n-2};
\\
\tilde\gamma_{12}^i &=& 0; 
\\
\alpha &=&  0;
\\
\beta^i &=& 0.
\end{eqnarray}
In terms of these $\gamma$ coefficients the $\Gamma_1\dots\Gamma_7$
are:
\begin{eqnarray}
\Gamma_1 &=& - 1;
\\
\Gamma_2 &=& +{1\over n-2};
\\
\Gamma_3 &=& - {1\over2(n-2)};
\\
\Gamma_4 &=& + \half;
\\
\Gamma_5^i &=& 0;
\\
\Gamma_6^i &=& 0;
\\
\Gamma_7^{ij} &=& +{n-3 \over 2(n-1)} \; [H_E]^{ij}.
\end{eqnarray}
When interpreted in terms of the effective Newton constant and related
quantities
\begin{eqnarray}
8\pi\;G_\effective &=&  
{\kappa_n^{-4}\over4} 
\left\{
{n-3\over n-2}\; \lambda 
\right\};
\\
8\pi\;G_\anomalous &=&  
0; 
\\
8\pi\;G_\quadratic &=& 
{\kappa_n^{-4}\over4}; 
\\
\Lambda_\effective &=&
{n-3 \over n-1} \;(\Lambda \; F^{-2/(n-2)} + \kappa_n^{-2} \; V_E)
\nonumber\\
&&
+ {\kappa_n^{-4}\over4} 
\Big\{ 
{n-3\over2(n-2)} \lambda^2 
-\kappa_n^2 \; {n-3\over2(n-1)} [H_E]^{ij}  \;(\L_\brane)'_i   \;(\L_\brane)'_j
\Big\}.
\end{eqnarray}
%

%-----------------------------------------------------------------------
\appendix
%-----------------------------------------------------------------------
\section*{Appendix C: Coefficients in the string frame}
%-----------------------------------------------------------------------
\label{A:coefficents-in-string}
%-----------------------------------------------------------------------
\setcounter{equation}{0}
%---------------------------------------------------------------------
\renewcommand{\theequation}{C.\arabic{equation}}
%-----------------------------------------------------------------------

String frame calculation: Consider $F(\phi)=\exp(-2\phi/\kappa_n)$,
$H(\phi)=-4\;\exp(-2\phi/\kappa_n)$. This corresponds to the dilaton
field in the string frame~\cite{generalized-junction}. For simplicity
the dilaton is assumed to be the only bulk moduli field. Then
\begin{eqnarray}
\E &=&  - {1\over n-2};
\\
H_E &=& {4\over n-2};
\\
\gamma_{11} &=& 0;
\\
\gamma_{12} &=& {n-1\over2};
\\
\gamma_{21} &=& \half;
\\
\gamma_{22} &=&  - {n-2\over4}.
\\
\tilde\gamma_{11} &=& 0;
\\
\tilde\gamma_{12}^i &=& \half; 
\\
\alpha &=&  -1;
\\
\beta^i &=& -{n-2\over2}.
\end{eqnarray}
In terms of these $\gamma$ coefficients the $\Gamma_1\dots\Gamma_7$
are:
\begin{eqnarray}
\Gamma_1 &=& - 1;
\\
\Gamma_2 &=& +{1\over n-2};
\\
\Gamma_3 &=& - {1\over(n-1)(n-2)};
\\
\Gamma_4 &=& + \half;
\\
\Gamma_5^i &=& 0;
\\
\Gamma_6^i &=& {n-3\over2(n-1)};
\\
\Gamma_7^{ij} &=& +{(n-3)(n-2) \over 8}.
\end{eqnarray}
%

%-----------------------------------------------------------------------------
%\clearpage
%------------------------------------------------------------------------------
 
%------------------------------------------------------------------------------

\begin{thebibliography}{99}%
%------------------------------------------------------------------------------
% SOME Spires codes included!
%-----------------------------------------------------------------------------
\bibitem{old-non-compact}
K.~Akama,
``An early proposal of `brane world' '',
[hep-th/0001113];
this is an electronic reprint of: ``Pregeometry'',
Lect.\ Notes Phys.\  {\bf 176} (1982) 267--271.
%%CITATION = HEP-TH 0001113;%%
\\
V.~A.~Rubakov and M.~E.~Shaposhnikov,
``Do We Live Inside A Domain Wall?'',
{\it Phys.  Lett.}\  {\bf B125} (1983) 136.
%%CITATION = PHLTA,B125,136;%%
\\
M.~Visser,
``An Exotic Class Of Kaluza-Klein Models'',
{\it Phys.  Lett.}\  {\bf B159}, 22 (1985).
[hep-th/9910093].
%%CITATION = HEP-TH 9910093;%%
\\
M.~Gell-Mann and B.~Zwiebach,
``Dimensional Reduction Of Space-Time Induced By Nonlinear Scalar Dynamics 
And Noncompact Extra Dimensions,''
Nucl.  Phys.\  {\bf B260}, 569 (1985).
%%CITATION = NUPHA,B260,569;%%
\\
E.~J.~Squires,
``Dimensional Reduction Caused By A Cosmological Constant,''
Phys.\ Lett.\  {\bf B167} (1986) 286.
%%CITATION = PHLTA,B167,286;%%
\\
P. Laguna--Castillo and R.~A. Matzner,
``Surfaces of discontinuity in five dimensional Kaluza--Klein models'',
Nucl. Phys. {\bf B282}, 542 (1987).
%%CITATION = NONE;%%
\\
G.~W.~Gibbons and D.~L.~Wiltshire,
``Space-Time As A Membrane In Higher Dimensions,''
Nucl.\ Phys.\  {\bf B287} (1987) 717.
%%CITATION = NUPHA,B287,717;%%
%------------------------------------------------------------------------------
\bibitem{large-compact}
I.~Antoniadis, 
``A Possible New Dimension At A Few TeV,''
{\it Phys.  Lett.}\ {\bf B246} (1990) 377.
%%CITATION = PHLTA,B246,377;%% 
\\
N.~Arkani-Hamed, S.~Dimopoulos and G.~Dvali,
``The Hierarchy Problem and New Dimensions at a Millimeter'',
{\it Phys.   Lett.}\ {\bf B429} (1998) 263-272.
[hep-ph/9803315].
%%CITATION = HEP-PH 9803315;%%
\\
``New dimensions at a millimeter to a Fermi and superstrings at a TeV,''
I.~Antoniadis, N.~Arkani-Hamed, S.~Dimopoulos and G.~Dvali,
{\it Phys.  Lett.}\ {\bf B436} (1998) 257. 
[hep-ph/9804398].
%%CITATION = HEP-PH 9804398;%%
\\
N.~Arkani-Hamed, S.~Dimopoulos and G.~Dvali,
``Phenomenology, Astrophysics and Cosmology of Theories with 
Sub-Millimeter Dimensions and TeV Scale Quantum Gravity''
{\it Phys.  Rev.} {\bf D59} (1999) 086004.
[hep-ph/9807344].
%%CITATION = HEP-PH 9807344;%%
\\
N.~Arkani-Hamed, S.~Dimopoulos and J.~March-Russell,
``Stabilization of Sub-Millimeter Dimensions: The New Guise 
of the Hierarchy Problem'',
[hep-th/9809124] 
%%CITATION = HEP-TH 9809124;%%
%------------------------------------------------------------------------------
\bibitem{RS}
L.~Randall and R.~Sundrum,
``A large mass hierarchy from a small extra dimension,''
{\it Phys.  Rev.  Lett.}\ {\bf 83} (1999) 3370. 
[hep-ph/9905221].
%%CITATION = HEP-PH 9905221;%%
\\
L.~Randall and R.~Sundrum,
``An alternative to compactification''
{\it Phys.  Rev.  Lett.}\ {\bf 83} (1999) 4690.
[hep-th/9906064].
%%CITATION = HEP-TH 9906064;%%
%-----------------------------------------------------------------------------
\bibitem{Gogberashvili}
M.~Gogberashvili,
``Four dimensionality in non-compact Kaluza-Klein model'',
Mod.  Phys.  Lett.\  {\bf A14}, 2025 (1999)
[hep-ph/9904383].
%%CITATION = HEP-PH 9904383;%%
\\
M.~Gogberashvili,
``Our world as an expanding shell'',
Europhys.  Lett.\  {\bf 49}, 396 (2000)
[hep-ph/9812365].
%%CITATION = HEP-PH 9812365;%%
\\
M.~Gogberashvili,
``Hierarchy problem in the shell-universe model'',\\{}
[hep-ph/9812296].
%%CITATION = HEP-PH 9812296;%%
\\
M.~Gogberashvili,
``Gravitational Trapping for Extended Extra Dimension'',\\{}
[hep-ph/9908347].
%%CITATION = HEP-PH 9908347;%%
%-----------------------------------------------------------------------------
\bibitem{Arkani-et-al}
N.~Arkani-Hamed, S.~Dimopoulos and G.~Dvali and N.~Kaloper
``Infinitely Large New Dimensions''
{\it Phys.  Rev.  Lett.}\ {\bf 84} (2000) 586-589.
[hep-th/9907209].
%%CITATION = HEP-TH 9907209;%%
%-----------------------------------------------------------------------------
\bibitem{Shiromizu}
T.~Shiromizu, K.~Maeda, and M.~Sasaki,
``The Einstein equations on the 3-brane world'',
[gr-qc/9910076].
%%CITATION = GR-QC 9910076;%%
%----------------------------------------------------------------------
\bibitem{Wands}
Ken-ichi Maeda and D. Wands,
``Dilaton gravity on the brane'',
[hep-th/0008188]
%%%CITATION = HEP-TH 0008188;%%
%----------------------------------------------------------------------
\bibitem{Mennim}
A. Mennim and R.A. Battye,
``Cosmological expansion on a dilatonic brane-world'',
[hep-th/0008192].
%%%CITATION = HEP-TH 0008192;%%
%----------------------------------------------------------------------
\bibitem{background-1}
M.~Sasaki, T.~Shiromizu and K.~Maeda,
``Gravity, stability and energy conservation on the Randall-Sundrum  brane-world,''
Phys.\ Rev.\  {\bf D62}, 024008 (2000)
[hep-th/9912233].
%%CITATION = HEP-TH 9912233;%%
\\
S.~Mukohyama, T.~Shiromizu and K.~Maeda,
``Global structure of exact cosmological solutions in the brane world,''
Phys.\ Rev.\  {\bf D62}, 024028 (2000)
[hep-th/9912287].
%%CITATION = HEP-TH 9912287;%%
%----------------------------------------------------------------------
\bibitem{background-2}
P.~F.~Gonz\'alez-D\'\i{}az,
``Cosmological predictions from the Misner brane,''
hep-th/0008193.
%%CITATION = HEP-TH 0008193;%%
\\
A.~Chamblin, H.~S.~Reall, H.~Shinkai and T.~Shiromizu,
``Charged brane-world black holes,''
hep-th/0008177.
%%CITATION = HEP-TH 0008177;%%
\\
C.~Cs\'aki, M.~L.~Graesser and G.~D.~Kribs,
``Radion dynamics and electroweak physics,''
hep-th/0008151.
%%CITATION = HEP-TH 0008151;%%
\\
A.~Mazumdar,
``Post-inflationary brane cosmology,''
hep-ph/0008087.
%%CITATION = HEP-PH 0008087;%%
\\
S.~Mukohyama,
``Quantum effects, brane tension and large hierarchy in the brane world,''
hep-th/0007239.
%%CITATION = HEP-TH 0007239;%%
\\
A.~Mazumdar,
``Interesting consequences of brane cosmology,''
hep-ph/0007269.
%%CITATION = HEP-PH 0007269;%%
\\
H.~Ishihara,
``Causality of the brane universe,''
gr-qc/0007070.
%%CITATION = GR-QC 0007070;%%
\\
P.~Bowcock, C.~Charmousis and R.~Gregory,
``General brane cosmologies and their global spacetime structure,''
hep-th/0007177.
%%CITATION = HEP-TH 0007177;%%
\\
S.~H.~Tye and I.~Wasserman,
``A brane world solution to the cosmological constant problem,''
hep-th/0006068.
%%CITATION = HEP-TH 0006068;%%
\\
D.~Langlois, R.~Maartens and D.~Wands,
``Gravitational waves from inflation on the brane,''
hep-th/0006007.
%%CITATION = HEP-TH 0006007;%%
\\
D.~Langlois,
``Brane cosmological perturbations,''
hep-th/0005025.
%%CITATION = HEP-TH 0005025;%%
\\
P.~F.~Gonz\'alez-D{\'\i}az,
``Misner-brane cosmology,''
Phys.\ Lett.\  {\bf B486}, 158 (2000)
[gr-qc/0004078].
%%CITATION = GR-QC 0004078;%%
\\
R.~Maartens,
``Cosmological dynamics on the brane,''
hep-th/0004166.
%%CITATION = HEP-TH 0004166;%%
\\
H.~Stoica, S.~H.~Tye and I.~Wasserman,
``Cosmology in the Randall-Sundrum brane world scenario,''
Phys.\ Lett.\  {\bf B482}, 205 (2000)
[hep-th/0004126].
%%CITATION = HEP-TH 0004126;%%
\\
N.~Deruelle and T.~Dolezel,
``Brane versus shell cosmologies in Einstein and Einstein-Gauss-Bonnet  theories,''
gr-qc/0004021.
%%CITATION = GR-QC 0004021;%%
\\
H.~Collins and B.~Holdom,
``Brane cosmologies without orbifolds,''
hep-ph/0003173.
%%CITATION = HEP-PH 0003173;%%
\\
H.~Shinkai and T.~Shiromizu,
``Fate of Kaluza-Klein bubble,''
Phys.\ Rev.\  {\bf D62}, 024010 (2000)
[hep-th/0003066].
%%CITATION = HEP-TH 0003066;%%
\\
M.~Cveti\v{c} and J.~Wang,
``Vacuum domain walls in D-dimensions: Local and global space-time  structure,''
Phys.\ Rev.\  {\bf D61}, 124020 (2000)
[hep-th/9912187].
%%CITATION = HEP-TH 9912187;%%
\\
J.~Garriga and T.~Tanaka,
%``Gravity in the brane-world,''
Phys.\ Rev.\ Lett.\  {\bf 84}, 2778 (2000)
[hep-th/9911055].
%%CITATION = HEP-TH 9911055;%%
\\
E.~E.~Flanagan, S.~H.~Tye and I.~Wasserman,
``Cosmological expansion in the Randall-Sundrum brane world scenario,''
Phys.\ Rev.\  {\bf D62}, 044039 (2000)
[hep-ph/9910498].
%%CITATION = HEP-PH 9910498;%%
\\
A.~Kehagias and E.~Kiritsis,
``Mirage cosmology,''
JHEP {\bf 9911}, 022 (1999)
[hep-th/9910174].
%%CITATION = HEP-TH 9910174;%%
\\
A.~Chamblin, M.~J.~Perry and H.~S.~Reall,
``Non-BPS D8-branes and dynamic domain walls in massive IIA  supergravities,''
JHEP {\bf 9909}, 014 (1999)
[hep-th/9908047].
%%CITATION = HEP-TH 9908047;%%
\\
H.A.~Chamblin and H.S.~Reall, 
``Dynamic Dilatonic Domain Walls'',
{\it Nucl.  Phys.}\ {\bf B562} (1999) 133-157
[hep-th/9903225].
%%CITATION = HEP-TH 9903225;%%
%----------------------------------------------------------------------
\bibitem{generalized-junction}
C.~{\Barcelo} and M.~Visser,
``Moduli fields and brane tensions: generalizing the junction conditions'',
Phys.\ Rev.\  {\bf D}, in press,
[gr-qc/0008008].
%%CITATION = GR-QC 0008008;%%
%----------------------------------------------------------------------
\bibitem{Israel-Lanczos-Sen}
%\bibitem{Israel}
W.~Israel,
``Singular hypersurfaces and thin shells in general relativity'',
{\em Nuovo Cimento}, {\bf 44B} [Series 10] (1966) 1--14; 
Errata---{\em ibid} {\bf 48B} [Series 10] (1967) 463--463.
%%CITATION = NUCIA,B44S10,1;%%
\\
%\bibitem{Lanczos}
K. Lanczos,
``Untersuching \"uber fl\"achenhafte verteiliung der materie in der 
Einsteinschen gravitationstheorie'', (1922), unpublished;
%%CITATION = NONE;%%
\\
Fl\"achenhafte verteiliung der materie in der Einsteinschen gravitationstheorie'', 
Ann. Phys. (Leipzig), {\bf 74}, (1924) 518--540.
%%CITATION = NONE%%
\\
%\bibitem{Sen}
N.~Sen, 
``\"Uber dei grenzbedingungen des schwerefeldes an unstetig keitsfl\"achen'',
Ann. Phys. (Leipzig), {\bf 73} (1924) 365--396.
%%CITATION = NONE;%%
%----------------------------------------------------------------------
\bibitem{conformal-frames}
V. Faraoni, E. Gunzig, and P. Nardone, 
``Conformal transformations in classical gravitational field theories
 and cosmology'', gr-qc/9811047.
\\
G.~Magnano and L.~M.~Sokolowski,
``On physical equivalence between nonlinear gravity theories 
and a general relativistic selfgravitating scalar field,''
Phys.\ Rev.\  {\bf D50}, 5039 (1994)
[gr-qc/9312008].
%%CITATION = GR-QC 9312008;%%
\\
D.~I.~Santiago and A.~S.~Silbergleit,
``On the energy-momentum tensor of the scalar field in scalar-tensor 
theories of gravity,''
Gen.\ Rel.\ Grav.\  {\bf 32}, 565 (2000)
[gr-qc/9904003].
%%CITATION = GR-QC 9904003;%%
\\
V.~Faraoni,
%``Illusions of general relativity in Brans-Dicke gravity,''
Phys.\ Rev.\  {\bf D59}, 084021 (1999)
[gr-qc/9902083].
%%CITATION = GR-QC 9902083;%%
%----------------------------------------------------------------------
\bibitem{scalars} 
C.~{\Barcelo} and M.~Visser,
``Traversable wormholes from massless conformally coupled scalar fields,''
Phys.\ Lett.\  {\bf B466} (1999) 127
[gr-qc/9908029].
%%CITATION = GR-QC 9908029;%%
\\
M.~Visser and C.~{\Barcelo},
``Energy conditions and their cosmological implications,''
gr-qc/0001099.
%%CITATION = GR-QC 0001099;%%
\\
C.~{\Barcelo} and M.~Visser,
``Scalar fields, energy conditions, and traversable wormholes,''
Class.\ Quant.\ Grav.\  {\bf 17} (2000) 3843
[gr-qc/0003025].
%%CITATION = GR-QC 0003025;%%
%----------------------------------------------------------------------
\bibitem{MTW}
C.~W.~Misner, K.~S.~Thorne, J.~A.~Wheeler,
{\emph{Gravitation}}, (Freeman, San Francisco, 1973).
%%%CITATION = NONE;%%
%----------------------------------------------------------------------
\bibitem{void}
C.~{\Barcelo} and M.~Visser,
``Brane surgery: Energy conditions, traversable wormholes, and voids,''
Nucl.\ Phys.\  {\bf B584} (2000) 415 [hep-th/0004022].
%%CITATION = HEP-TH 0004022;%%
%-----------------------------------------------------------------------------
\bibitem{edge}
C.~{\Barcelo} and M.~Visser,
``Living on the edge: Cosmology on the boundary of anti-de Sitter space,''
Phys.\ Lett.\  {\bf B482} (2000) 183
[hep-th/0004056].
%%CITATION = HEP-TH 0004056;%%
%----------------------------------------------------------------------
\bibitem{surgery}
M.~Visser,
``Traversable Wormholes From Surgically Modified Schwarzschild Space-Times,''
Nucl.\ Phys.\  {\bf B328} (1989) 203.
%%CITATION = NUPHA,B328,203;%%
%----------------------------------------------------------------------
\bibitem{book}
M.~Visser,
{\sl Lorentzian Wormholes: from Einstein to Hawking},
(AIP Press, New York, 1995).
%%%CITATION = NONE;%%
%----------------------------------------------------------------------
\end{thebibliography}
\end{document}